\documentclass[11pt,a4paper]{article}
\usepackage{amsmath}
\usepackage{amsfonts}
\usepackage{amssymb,mathrsfs}
\usepackage{t1enc}
\usepackage[latin1]{inputenc}
\usepackage[english]{babel}
\usepackage[numbers]{natbib}
\usepackage{amssymb}
\usepackage[final]{pdfpages}
\usepackage{caption}
\usepackage{url}

\DeclareMathOperator{\arccosh}{arcCosh}

\usepackage[colorlinks=false,linktocpage=true,backref,pagebackref]{hyperref}
\hypersetup{allbordercolors=[rgb]{0,1,0}}
\hypersetup{allbordercolors=green}

\newcommand{\R}{\mathbb R}

\newcommand{\beq}{\begin{equation}}
\newcommand{\eeq}{\end{equation}}
\newcommand{\beqarr}{\begin{eqnarray}}
\newcommand{\eeqarr}{\end{eqnarray}}
\newcommand{\beqa}{\begin{eqnarray*}}
\newcommand{\eeqa}{\end{eqnarray*}}
\begin{document}
\thispagestyle{empty}

\title{\bf \large A Weyl geometric  scalar field approach to the dark sector}
\author{\normalsize Erhard Scholz\footnote{University of Wuppertal, Faculty of  Math./Natural Sciences, and Interdisciplinary Centre for History and Philosophy of Science, \quad  scholz@math.uni-wuppertal.de}}
 \date{\small \today} 
\maketitle
\begin{abstract}
This paper explores the  dark sector (dark matter and dark energy) from the perspective of Weyl geometric scalar tensor theory (integrable Weyl geometry). In order to account for the galactic dynamics  successfully modelled  by  MOND (``modified Newtonian dynamics''), the non-minimally coupled scalar field  considered here has  a Lagrangian with two non-conventional contributions in addition to  a standard kinetic term: one is  inspired by Bekenstein/Milgrom's RAQUAL (``relativistic a-quadratic Lagrangian'') from  1983, the other one by a second order term introduced in cosmological studies  by  Novello et al. in 1993. {\bf See, however, the error warning below}.
We consider the transition to the Einstein gravity on one hand and to  scalar field cosmology in the FRW framework on the other.  A  bouncing  cosmological model  is tentatively discussed at the end.
\end{abstract}

\section*{Error warning}
\addcontentsline{toc}{section}{\protect\numberline{}Error warning}
{\bf Equ. (\ref{eq Theta braz}) is wrong; it does not take  contributions to  $\frac{\delta L^{braz}}{g^{\mu\nu}}$ into account, which are  due to   the covariant derivative of the (scale covariant) gradient of the scalar field $\phi$. Therefore the contribution of $L^{braz}$ to the  energy tensor of the scalar field is wrong. This  flaw is fatal for the  derivations in the Milgrom regime; the whole argumentation of part 2 can  no longer be upheld. It is here reproduced for documentary reasons only.
}

\setcounter{tocdepth}{2}
\tableofcontents

\section*{Introduction}
\addcontentsline{toc}{section}{\protect\numberline{}Introduction}
Shortly after the  rise of  broader interest in dark matter at the turn to the 1980s Bekenstein and Milgrom explored a  Lagrangian approach for explaining  the flat rotation curves of galaxies in modified gravity  (MOND)   \citep{Bekenstein/Milgrom:1984,Bekenstein:2004,Milgrom:2014}. They  assumed a scalar field $\phi$ with an unconventional ``a-quadratic''  kinetic Lagrangian term (AQUAL)  and indicated  how, at least in principle,  a relativistic version called RAQUAL can be formulated  along these lines.  The authors observed that this approach did not provide for  gravitational refraction  of light sufficiently exceeding that of baryonic matter. In consequence a series of other, more or less ad-hoc, modifications of Einstein gravity  were proposed for  addressing this problem. A generic feature  of them is  the idea of a cubic kinetic term  (enveloped in  a more general transition function to Einstein gravity) which reproduces the achievements of MOND at the galactic level. Usually other  fields are added, which take care of the necessary gravitational light refraction. Among them there are  tensor-vector scalar field theory (TeVeS), generalized Einstein-Aether theory (gEA),  superfluid theory, conformal emergent gravity  \citep{Bekenstein/Milgrom:1984,Bekenstein:2004,Sanders:2007,Berezhiani/Khoury:2016,Hossenfelder:2017,Hossenfelder/Mistele:2018,
Skordis/Zlosnik:2019}, to name the best known ones; most recently a ``new relativistic theory for modified Newtonian dynamics'' (nrMOND) \citep{Skordis/Zlsonik:2021}. For expositions and comparisons of some of the  approaches with this background see \citep{Skordis/Zlosnik:2011,Famaey/McGaugh:MOND,Milgrom:2014}. Modified gravity theories in the cosmological range have a different motivation; for reviews see \cite{Clifton/Ferreira_ea:2012,Faraoni:2004,Capozziello/Faraoni}.

In many of these  approaches new hypothetical  degrees of freedom of the gravitational field  are  introduced  more or less artificially in order to  emulate the effects which in the  $\Lambda$CDM  approach are considered as the effects of dark matter. 
The present proposal  is  more parsimonious. It  shows  that {\em  one scalar field and}  a well founded  {generalization of the metric} from Riemannian to  {\em  (integrable) Weyl geometry} (IWG) suffice for explaining dark matter effects at the astrophysical level of galaxies (and probably also of galaxy clusters).  The corresponding gravitational light bending  is induced by a scalar field the  kinetic term of which is  complemented, besides a  cubic (RAQUAL inspired) one,  by a second order derivative term  first proposed in \citep{Novello/Oliveira_ea:1993} in a cosmological context. This term endows  the scalar field with an energy-stress tensor which represents a kind of dark matter/energy {\em sui generis}, while the scalar field  modifies gravity similar to Jordan-Brans-Dicke theory (JBD) \citep{Fujii/Maeda}. Here the latter is  related to the (integrable) scale connection of Weyl geometry;  
the present approach will therefore be called a {\em Weyl geometric ``dark'' scalar theory} (WdST). 
 
 The paper starts with a short discussion of how the conformal transformations in  JBD theory can be expressed in terms of Weyl geometry.  The scale covariant scalar field of JBD is then  easily taken over to  IWG. In this framework the  different frames of JBD  appear as the different scale gauges of one and the same integrable  Weylian metric. This results in a slight but important  shift in the perspective on  free fall trajectories of test bodies and on the  question which affine connection ought  to be considered as ``physical'' (sec. \ref{section JBD-IWG}). 
 
 If one wants to  explain MOND-type galactic dynamics in the framework of a  relativistic  scalar tensor theory, one  needs a  more general Lagrangian than in JBD.  Moreover, although Einstein gravity will continue to hold in large regions of the universe, 
 at  the galactic and cluster level, i.e.,  in very weak field regions, it has to be changed,  just like Newton theory is being changed by MOND.  There exist many, often quite  different  reasons for looking at alternatives to Einstein gravity  at the cosmological level. It is  no longer clear that one universal Lagrangian of a new ``fundamental'' theory  will cover  all these gravitational regimes. More cautiously the next section  proposes  to  distinguish between  different gravitational regimes (\ref{subsection grav regimes 1}). The main part  of the section  develops  a Weyl geometric scalar field approach to galactic and cluster dynamics, a region  here called the {\em Milgrom regime}  of WdST.  
 
 After introducing the Lagrangian of the Milgrom regime  at the beginning of sec.~\ref{subsection Milgrom regime} the dynamical equations for the Riemannian component of the Weylian metric (generalized Einstein equation) and the scalar field equation (Milgrom equation) are derived, with some technicalities shifted to  the appendix. A short look at the energy-stress tensor of the scalar field and the contribution of the scalar field to the (coordinate) acceleration in the Einstein gauge follow. While  an investigation of the static weak field approximation  leads to a Newton-type approximation for the $(0,0)$-component of the metric sourced by baryonic matter only, the spatial components  are strongly  influenced by the scalar field pressures. This has the effect   that the gravitational light refraction stands in agreement with the modification of the particle acceleration by the scalar field (sec.~\ref{subsection weak field approximation}). 
Another upshot is the observation that  a special type of MOND dynamics arises in the flat space limit. The effects for  gravitational light bending can be studied by   relativistic corrections to the latter (sec.~\ref{subsection flat space limit}).

We then come back to the question of how to represent  other gravitational regimes mentioned above  in WdST  (sec.~\ref{section  other regimes}). Although it is clear how to incorporate Einstein gravity as a special  case of WdST mathematically (by trivializing the scalar field dynamics), the physical reasons for the transition remain open.  Also open is the question of how to represent the cosmological regime in WdST, even under the  idealising symmetry assumption of FRW geometry. A first exploration of the question, including a new  bouncing cosmological model is given in sec.~\ref{subsection FRW regime}. It is followed by a r\'esum\'e and  a discussion of what has been achieved in the paper (sec.~\ref{section Outlook}). Some technical derivations are dealt with in the appendix (sec.~\ref{section appendix})

\section{JBD and Weyl geometric scalar tensor theory \label{section JBD-IWG}}

\subsection{\small A reminder on JBD gravity }
Jordan-Brans-Dicke theory (JBD) assumes a scalar field $\Phi$ contributing to the gravitational Lagrangian (Hilbert term)  in the well known way:
\beq L_J= \frac{1}{2}\Phi R - \frac{\omega}{\Phi}\,\partial_{\nu}\Phi \partial^{\nu}\Phi + L_{m}\, , \qquad \mathcal{L}_J = L_J \sqrt{|\mathit{det}g|}\vspace{0em} \,  \label{eq Jordan frame}
\eeq
Here $R$ denotes the  Riemannian scalar curvature of a metric $g_J= (g_{\mu\nu})$ with signature $(-+++)$ in the {\em Jordan frame}. 

Dicke postulated 
 that the  basic laws of physics ought to be  independent of the choice of measuring units  \cite[p. 2163]{Dicke:1962}. The acceptance of this postulate would  demand a theory which is invariant under  (nonsingular) conformal transformations, i.e.~scale transformations, among different  {\em frames}. That was  similar to  Weyl's  proposal (1918) of a scale gauge geometry \citep{Weyl:GuE_English}, if one  includes conformal rescaling of fields.\footnote{Weyl scaling is based on conformal rescaling of the metric $g(x) \mapsto \tilde{g}(x)= \Omega(x)^2 g(x)$i.e., length and time intervals are  rescaled by the point-dependent factor $\Omega(x)$. The speed of light $c$ is scale invariant  and the Planck constant $\hbar$ is postulated as such, i.e., energy and mass scale by the factor $\Omega^{-1}$. Respecting the new SI conventions, it is clear how dimensional  quantities and thus also fields have to be rescaled. This is an extension of Weyl's proposal  and will be called {\em Weyl scaling}. It is logically equivalent to the rescaling in high energy physics, although with inverted scale weights. \label{fn Weyl scaling}} 
 
  Weyl generalized the Riemannian metric to  what was later called a {\em Weylian metric}. The latter may be expressed in different scale covariantly connected forms, so-called (scale-)gauges. They are characterized by pairs $(g,\varphi)$ consisting of a Riemannian  metric $g$ (of any signature) and a differential 1-form $\varphi$, the Weylian scale connection (in the  physical literature  often called the ``Weyl field''). Different gauges of the Weylian metric, $(g,\varphi)$ and $(\tilde{g}, \tilde{\varphi})$ are connected by the rescaling of the {\em Riemannian components} $\tilde{g}=\Omega^2 g$ and an accompanying gauge transformation of the connection $\tilde{\varphi}= \varphi- d\log \Omega$ \citep{Blagojevic:Gravitation,Ohanian:2016,Dengiz:2017}.\footnote{For a short introduction see, e.g.,  \citep[sec. 11.2]{Scholz:2018Resurgence} and the literature given there in  footnote 7.}
 
 JBD theory  may  be rephrased in terms of   {\em integrable  Weyl} geometry (IWG) \citep{Quiros_ea:2013,Almeida/Pucheu:2014,Scholz:Paving}. Integrable means here that the scale connection $\varphi$ of any gauge $(g,\varphi)$ of the Weylian metric is {\em pure gauge}, 
 \beq \varphi= - d \log \Omega \, , \label{eq varphi Einstein gauge 1}
 \eeq
  where $\Omega$ is the rescaling function s.th. $g= \Omega^2 g_J$.  Of course, different {\em scale gauges} in IWG correspond to different frames in JBD. 
Here we count the scaling weight  of the scalar field  as $w(\Phi)=-2$ according to  $w(g_{\mu\nu})=2$, which means that length quantities scale with  weight 1.\footnote{By obvious reasons elementary particle physics prefers inverted signs of the weight convention.}
As the Jordan frame expresses the Weylian metric in Riemannian terms we speak of it as the {\em Riemann gauge} of the underlying IWG. 

 Empirical interpretation of observable quantities presupposes a particular choice of measuring units, which results in an appropriate gauge fixing. Observable quantities are thus expressed in a  specified frame/gauge which we will call the {\em measuring gauge}. The measuring gauge of a scale covariant field $X$  will be denoted by $X_m$. Of course one may want to know the reason for such a choice, at best given by a ``breaking mechanism'' of the scale symmetry. At the moment we leave this question open, but come back to it in section \ref{section Outlook} ($\rightarrow$ Higgs gauge). 
  Because of the preferred relation to Einstein gravity  the scale gauge in which the scalar field is a constant, $\phi_0$, it is called a scalar field ($\phi$-)gauge. If the constant is normed to  the gravitational constant
  \beq
   \phi_0^2 =(8\pi G_N [c^{-4}])^{-1} = (8 \pi \varkappa)^{-1} \, ,  \label{eq Einstein gauge}
   \eeq 
($G_N$ the Newton constant and $\varkappa=G_Nc^{-4}$)
    it is  called the
{\em Einstein gauge}, respectively frame;\footnote{Conventions for (physical) dimensions here: $[x_{\mu}]=1$ (dimensionless coordinates),  $[g_{\mu \nu}]=L^2$, 
then $[\varkappa]=L E^{-1}$ ($L$ length, $E$ energy etc.)}
  a field $X$ in this gauge will be  denoted by $X_E$. More gauges will be introduced  in the following.  

The Levi-Civita connections of the metrics in different frames define inequivalent affine structures. So an important question in JBD is:  
Which affine connection governs  the inertial structure of JBD gravity?  (Here the range of ``affine connections'' is  implicitly delimited by the 
Levi-Civita connections of all frames conformal to $g$.)  The debate often reduces the question to the decision between the Jordan and Einstein frames \citep[sec. 2.1]{Faraoni:2004}, \citep[sec. 3]{Fujii/Maeda}. From the point of view of IWG one would pose the question differently, as we discuss in the next subsection (compare \citep{Quiros_ea:2013,Almeida/Pucheu:2014}).

\subsection{\small The point of view of  integrable Weyl geometry (IWG)}
Different from JBD theory, in  Weyl geometry exists a  {\em uniquely determined affine connection}  compatible with the Weylian metric. Given any gauge $(g,\varphi)$ it can be written as
\beq \Gamma(g,\varphi)= \Gamma(g) + \Gamma(\varphi) \, ,  \label{eq Gamma(g,varphi)}
\eeq
with $ \Gamma(g) $ the Levi-Civita connection of $g$ and 
\beq \Gamma(\varphi)^{\mu}_{\nu \lambda} = \delta^{\mu}_{\nu}\varphi_{\lambda} + \delta^{\mu}_{\lambda}\varphi_{\nu} - g_{\nu \lambda}\varphi^{\mu} \label{eq Gamma(varphi)}
\eeq 
  the contribution of the scale connection. 
$\Gamma(g,\varphi)$  induces a well determined (unique) scale covariant differential $D_{\mu}$ of fields, and leads to a generalized Riemannian curvature of Weyl geometry $\mathit{Riem}(g,\varphi)$ which is no longer antisymmetric in the first two entries. The resulting Ricci curvature $\mathit{Ric}(g,\varphi)$ is scale invariant, while the scalar curvature $R(g,\varphi)$ is scale covariant of weight $w(R)=-2$, 
\beq  R(g,\varphi)= R(g) - (n-1)(n-2)\varphi_{\nu}\varphi^{\nu}- 2(n-1) \nabla(g)_{\nu}\varphi^{\nu}\, , \label{eq R(g,varphi)}
\eeq 
where $R(g)$ denotes the Riemannian scalar curvature of $g$, $\nabla(g)$ the covariant derivative of $\Gamma(g)$, and $n$ is the dimension  of the Weylian manifold,  
see \citep{Ghilencea:2019JHEP,Yuan/Huang:2013}, also \citep[sec. 15.2]{Adler/Bazin/Schiffer}.

Of course the path structure of  affine geodesics $\gamma(\lambda)$, the autoparallels with regard to (\ref{eq Gamma(g,varphi)}), is independent of the chosen gauge; only the parametrization may change. We chose it such that in any gauge for non-null geodesics we get  $g(\dot{\gamma},\dot{\gamma})= \pm 1$; i.e.~we use scale dependent parametrizations of geodesics with weight $w(\dot{\gamma})=-1$.

It is  known  that a general,  non-integrable scale connection (a ``Weyl field''~$\varphi$ with $d\varphi\neq	 0$, i.e., non-vanishing field strength), leads to a mass term close to the order of Planck energy  \citep{Smolin:1979,Drechsler/Tann,Cheng:1988}. Even if it were physical,  it could be integrated out  at the energy scales of classical field theory, which we coinsider here.  This leaves only the scalar degree of freedom  for the integrable case. 
The arising 
{\em  Weyl geometric scalar tensor}  theory  (WST) can also be arrived at by a slight modification of JBD gravity; then the scale connection $\varphi$ is integrable from the outset. 
It has no own dynamical degree of freedom of its own but ``shares'' it, so to speak with the scalar field. 

Its  Lagrangian has  no dynamical term of the Weyl field
 and is of a form close to JBD:
\beq L= L^{(H)} +  L^{(kin)} + L^{(V)}  + L^{(m)}\, , \qquad \mathcal{L} = L \sqrt{|\mathit{det}g|} \, , \label{eq Lagrangian WST} 
\eeq
where all building blocks $\mathcal{L}^{(X)}$ are scale invariant. The gravitational part is a Hilbert term, 
\beq
\mathcal{L}^{(H)}=\frac{1}{2}\phi^2  R(g,\varphi) \, , \label{eq L^H}
\eeq
with $R(g,\varphi)$ the scalar curvature of Weyl geometry (weight  -2), 
$ \phi$  a gravitational scalar field  (weight -1) similar to JBD theory ($\Phi=\phi^2$).  $L^{(kin)}$ and $L^{(V)}= - V(\phi)$, respectively their densities, are the kinetic and potential terms of the scalar field; $L^{(m)}$ is the matter term brought into a scale covariant form (weight -4) (cf.  sec. \ref{subsection FRW regime}). The simplest choice for the potential is the quartic monomial
\beq V(\phi)= \frac{\lambda_4}{4}\phi^4  = V_4(\phi) \label{eq L V-4} \, ;
\eeq 
a  more refined one  (biquadratically coupled to the Higgs field)  will be discussed in sec.\ref{section Outlook}.  
It is  known  that a general,  non-integrable scale connection  leads to a mass term close to the order of Planck energy  \citep{Smolin:1979,Drechsler/Tann,Cheng:1988}. Even if it were physical,  it could be ``integrated out''  at the energy scales of classical field theory considered here. This leaves only the scalar degree of freedom  for the integrable case.

 
For a quadratic kinetic term  $L^{(kin)}$ of the form 
\beq \quad L^{(q-kin)}=- \frac{\alpha}{2} D_{\nu}\phi D^{\nu}\phi \, \,  \label{eq L q-kin}
\eeq
the Lagrangian (\ref{eq Lagrangian WST}) is essentially the one of JBD theory, written in scale covariant form. Remember that $D$ denotes the scale covariant derivative, $D_{\mu}\phi=\partial \phi - \varphi_{\mu}\phi$. 
For a scalar field with  $D_{\mu}\phi=0$ (vanishing scale covariant gradient) the  Einstein gauge coincides with the Riemann gauge and IWG gravity  reduces to Einstein gravity.
Let us add that for the Weyl geometric dark scalar tensor theory studied here  (WdST) we will add two more terms (\ref{eq L-Dphi3}), (\ref{eq L-braz}) to $L^{(kin)}$.\footnote{The Palatini variation approach used in \citep{Poulis/Salim:2011,Pucheu_ea:2016}  and other work of the Brazilian group implies the constraint {\em  Einstein gauge = Riemann gauge}. The approach, there  called {\em Weyl integrable spacetime} (WIST),  is thus a  very special case of IWG gravity. Even for $ L^{(kin)}= L^{(q-kin)}$  it boils down to a scale covariant description of Einstein gravity, cf.  \citep[sec. 4.1.C]{Duerr:2021}, \citep{Ghilencea:2021}.}

 Equations which  hold only in Einstein gauge (frame) will be denoted  by $ \underset{E}{\doteq}$, similarly for the other gauges. For example   $L^{(H)} \underset{E}{\doteq} \frac{1}{2}\phi_0^2\, R$, with  $\phi_0^2 = (8 \pi \varkappa)^{-1}$,   thus $\phi_0 \sim E_{Pl}$ (reduced Planck energy). 
Let us write the scalar field in Riemann gauge/Jordan frame in an exponential form
\beq \phi(x) \underset{R}{\doteq} \phi_R(x)=\phi_o e^{-\sigma(x)} \, .
\eeq 
\noindent
In Einstein gauge $\phi_E$ is constant  due to (\ref{eq varphi Einstein gauge 1});  the scale connection is 
\beq \varphi_E=d\sigma\, , \qquad \mbox{respectively} \quad \varphi_{\mu} \underset{E} \doteq \partial_{\mu} \sigma \, . \label{eq varphi Einstein gauge 2}
\eeq 
 Thus   the kinetic term  reduces  to 
$ L^{(q-kin)}  \underset{E} \doteq -\frac{\alpha}{2} \phi_0^2\, \partial_{\nu}\sigma  \partial^{\nu}\sigma \,  
$
 and the Lagrangian to 
\beq L\underset{E}\doteq  (8 \pi \varkappa)^{-1}\Big( \frac{1}{2}  R(g,d\sigma) - \frac{\alpha}{2} \partial_{\nu}\sigma  \partial^{\nu} \sigma -  \Lambda \Big) \quad  + L^{(m)}\, ,\label{eq Lagrangian Eg}
\eeq 
with $\Lambda = \lambda_4 \,\phi_0^2$, where $\lambda_4$ is ``cosmologically'' small, i.e.~it contains a very small ``hierarchy'' factor,  $\lambda_4 =(\lambda \beta^{-1})^2$, where  $\beta^{-1}$ bridges the gap between the Planck scale and the scale of cosmologically small quantities like $\Lambda^{\frac{1}{2}}$ or $a_0 \hbar$, with $a_0$ the MOND constant introduced below (\ref{eq a_0}).

At first glance this  resembles a Lagrangian of a minimally coupled field with a  cosmological constant, but it is not, as the Hilbert term is formed from the Weyl geometric scalar curvature $R(g,d\sigma)$. The ``coupling'' of $\sigma$ to the Hilbert term, contained in (\ref{eq L^H}) has  taken on a specific form.
The energy momentum tensor $\Theta = (8 \pi \varkappa) T^{(\phi)}$
of the scalar field and the scalar field equation come  out differently from  the analogous Lagrangian in a Riemannian framework. From \citep{Smolin:1979,Blagojevic:Gravitation,Drechsler/Tann,Scholz:2020GRG} we know:
\beq \Theta_{\mu\nu} \underset{E}{\doteq} (\alpha + 6)\, \partial_{\mu}\sigma \partial_{\nu}\sigma - \big[ (\frac{\alpha}{2}+3)\, \partial_{\lambda}\sigma \partial^{\lambda}\sigma\, - \lambda \phi^2 \big]g_{\mu\nu} \label{eq Theta}
\eeq 
\beq \nabla(g)^2 \,\sigma \underset{E}{\doteq} \nabla\hspace{-0.2em}(g)_{\hspace{-0.1em}\lambda}\partial^{\lambda}\sigma = \frac{8\pi \varkappa}{3+\frac{\alpha}{2}} (\rho^{(bar)}-3 p^{(bar)})\,  \label{eq scalar field D2}
\eeq 
 $\nabla\hspace{-0.2em}(g)$ denotes  the covariant differentiation with respect to the Levi-Civita connection of $g$ etc. 
The values 6 and 3 depend on the dimension $n$ of the spacetime, here $n=4$.
 With minimal coupling of $\sigma$ (respectively $\phi$) no  matter coupling  would arise in  (\ref{eq scalar field D2}).

In the JBD tradition an equivalent result is derived  in several steps: variation in the Jordan frame, conformal transformation to the Einstein frame, and a field redefinition of the scalar field \citep{Fujii/Maeda}. The result is equivalent to our eg.~(\ref{eq Lagrangian Eg}).

 (\ref{eq Theta}) shows a  decomposition  $\Theta = \Theta^{(de)}+\Theta^{(dm)}$,  in a term proportional to $g$ which may be considered as ``dark energy'', $\Theta^{(de)}$,  and another one, $\Theta^{(dm)}$, which  
 may be considered as dark matter-like, if one uses a very generous concept of ``dark matter'', 
\[ \Theta_{\mu \nu}^{(dm)} =  (\alpha + 6)\, \partial_{\mu}\sigma \partial_{\nu}\sigma\, , \qquad  \Theta^{(de)}= - \tilde{\Lambda}(x)g \, ,
\]
where  $\tilde{\Lambda}(x) \underset{E}{\doteq}(\frac{\alpha}{2}+3)\,  \partial_{\lambda}\sigma(x) \partial^{\lambda}\sigma(x) + \Lambda $ and $\Lambda$ like in  (\ref{eq Lagrangian Eg}. 
A similar decomposition holds for any scalar field theory  (by the way, also in the case of  minimal coupling).

\subsection{\small Free fall  and  light bending  in JBD and IWG gravity \label{subsection Differences IWG JBD}}
JBD and IWG gravity make different assumptions for the dynamics of test particles. 
\noindent
a)  {\em JBD gravity} assumes  coupling of matter to the metric $g$ of one of the  frames, usually to the Jordan frame or to the Einstein one. Freely falling particles then follow the trajectories of the Levi-Civita connection $\Gamma(g)$ of the respective frame. \\
\noindent
b) {\em Weyl geometric} scalar tensor gravity, as it is understood here,  assumes a matter Lagrangian $\mathcal{L}^{(m)}$ in a scale invariant form.
Freely falling matter particles then  follow the geodesics of the invariant Weyl geometric affine connection  $\Gamma(g,\varphi)$ which  includes  terms in the scale connection. The latter express an additional acceleration which adds to the one induced by the Levi-Civita connection of the Riemannian component $g$ (see sec. \ref{subsection acceleration} and \citep[secs. 2.5,  6.3]{Scholz:2020GRG}). 

The electromagnetic field and null geodesics depends only on the conformal structure; the gravitational bending of light can be described in any gauge  by the respective Riemannian component $g$ of the Weylian metric; it does not depend on (``respond to'') the the respective scale connection, while matter particles do. The  option b) may thus give the impression that  inertial motion  of particles and light rays follow different ``laws''.
 But this is not the case, both are governed by one and the same geodesic structure of the Weyl metric.  This may give rise to problems in the weak field approximation, but it  need not do so.
 
In  sec.~\ref{subsection weak field approximation} it will be shown that the impact of the stress components of the scalar field energy tensor in the present approach  are  strong enough to induce a non-negligible contribution to gravitational light bending, which is consistent with the additional acceleration for particles due to the scale connection in the Einstein frame. For this result it is important to choose  a more elaborate kinetic term for the scalar field than the standard one (\ref{eq L q-kin}).

\subsection{\small Coordinate acceleration  in IWG  \label{subsection acceleration} }
If  the matter Lagrangian $\mathcal{L}^{(m)}$ is written   in a scale invariant form,\footnote{This is the case for the SM fields of high energy physics before introducing the matter term of the Higgs field.  For classical matter fields the scale  invariance of   $\mathcal{L}^{(m)}$ can be introduced formally. Although this  looks artificial at the first sight, it may just as well express a deeper truth resulting from decoherence of the quantum domain.}
 the  energy momentum tensor $T^{(m)}$  scales with weight $w(T^{(m)})=-2$.
A Geroch-Jang type argument \citep{Geroch/Jang:1975} shows  that  in such a  framework test bodies  move along  timelike geodesics of integrable Weyl geometry \citep[app. 6.3]{Scholz:2020GRG}. In the perspective of JBD this would amount to coupling of matter to the Jordan frame metric. From the point of view of IWG, however, {\em  matter does not  couple to the metric of  a specific frame} but to the  scale invariant affine connection of Weyl geometry.

For low velocity trajectories parametrized in proper time $\tau$ the  coordinate  acceleration of Riemannian geometry is 
\beq  a^{j} = - \Gamma^j_{00}  \label{eq a^j}
\eeq
  \citep[p. 153f.]{Carroll:Spacetime} or \citep[eg.~9.1.2]{Weinberg:Cosmology_1972}. This also holds in the Weyl geometric context.
 (\ref{eq Gamma(g,varphi)}) shows that  the coordinate acceleration has  two contributions 
\beq
a^{j} = a(g)^{j}+ a(\varphi)^{j}\,      \qquad (j=1,2,3). \label{eq total acceleration}
\eeq
The first one, $a(g)^{j}=- \Gamma(g)^j_{00}$, is due to the Levi Civita of the Riemannian component. The second one derives from the scale connection, $  a(\varphi)^{j}= - \Gamma(\varphi)^j_{00}$; it vanishes only in the Riemann gauge.  
In Einstein gauge the scale connection is determined by the scalar field $\phi$, respectively $\sigma$ ($\varphi=d\sigma$); we therefore  denote the additional acceleration  $a(\varphi)$ also by $a(\sigma)$. In any scale gauge $(g,\varphi)$ different from the Riemannian one ($\varphi  \neq 0$),  the scale connection distracts  free fall  away from the Levi-Civita trajectories of the respective Riemannian component $g$.
This general property of IWG is important 
for the   dark scalar field theory of galactic and cluster dynamics which we now turn to.

 \section{Weyl geometric dark scalar field theory  (WdST) \label{section WdST} } 

\subsection{\small  Gravitational regimes of WdST \label{subsection grav regimes 1}}
The experience with MOND 
and its relativistic generalizations \citep{Milgrom:2014,Skordis/Zlsonik:2021}, \citep[sec. 5]{McGaugh:2021}  is here  taken into account by assuming  three different regimes  distinguished by (the norm of) the gradient  of the scalar field $|\nabla \phi|$, respectively $|\nabla \sigma|$. The gravitational dynamics is governed by different, but related  Lagrangians:
\begin{itemize}
\item[(i)] For $\nabla \sigma=0$ we are in the ordinary gravity regime governed by the dynamics of Einstein gravity (EG).  It has  been studied for more than a century in  detail and with great success. As  a special case of  IWG gravity it will here be considered   as
 the {\em Einstein regime} of WdST.
\item[(ii)] For spacelike gradient and $|\nabla \sigma|$  close to the order of the MOND constant $a_0 \approx \frac{H_0}{6}$  (see \ref{subsection 3 regimes and transition}) we are in an  (ultra) weak gravity regime typically obtained at the level of outer galaxies and clusters. In  WdST its dynamics is characterized by a scale covariant scalar field which modifies   Einstein gravity. The weak field approximation results in a specific type of MOND dynamics.\footnote{The specific type is characterized by the MOND-typical ``interpolation function'' $\nu (y)= 1+y^{-\frac{1}{2}}$ valid  in the Milgrom regime, see sec \ref{subsection flat space limit}.}
 This is the motivation for  calling it the {\em Milgrom regime} of WdST. In this regime the scalar field is  governed by two untypical kinetic terms: 
a cubic kinetic term similar to the one studied by Bekenstein/Milgrom  in the deep MOND case \citep{Bekenstein/Milgrom:1984}; it leads to a scale covariant relativistic Milgrom equation for the scalar field (generalizing the non-linear Poisson equation of ordinary MOND). Moreover  a second order kinetic term is assumed; it was first introduced in a cosmological context by Novello et al.  \citep{Novello/Oliveira_ea:1993}) and endows the scalar field with  non-negligible energy momentum important for gravitational light refraction. 
\item[iii)] For $\nabla \sigma$ timelike (or  for an extremely small norm in the  spacelike case;  cf. sec. \ref{subsection 3 regimes and transition}) we are in a regime in which, under the idealizing assumption of  homogeneity/isotropy, large scale models of  FRW type are informative. It will be called the {\em FRW regime} of cosmology.  The present experience with the WdST approach  indicates that  a straight-forward extrapolation of the  dynamics of the Milgrom regime to this scale is  unrealistic (sec. \ref{subsection FRW regime}).  Weyl geometric or conformal approaches to the FRW regime are shortly  discussed, but   the research in this respect is far from conclusive.
\end{itemize}


In  MOND   and some of its relativistic generalizations a free function $f(X)$ with appropriate asymptotic, or long range  behaviour is used for characterizing the transition between the MOND/Milgrom and the Newton/Einstein regimes \citep{Bekenstein:2004,Skordis/Zlosnik:2011}. In the present approach the passage from one regime to the other will   be  described by means of smooth transition functions between the different Lagrangians (sec. \ref{subsection 3 regimes and transition}). 
Berezhiani/Khoury have proposed that the  physical cause for the  transition  between the Einstein regime and the Milgrom regime 
 may lie in a phase transition of   a hypothetical substrate to a  superfluid state \citep{Berezhiani/Khoury:2015,Berezhiani/Khoury:2016}. If successful, this approach may also lead to an explanation of the appearance of fractional powers in the kinetic term of the scalar field (see below); but at the moment this seems still unclear. 
 
Here we 
 start by investigating  the dynamics in the Milgrom regime.  A short  discussion of the other regimes and some remarks on the  transition   follow in sec. \ref{subsection 3 regimes and transition}. 

\subsection{\small Milgrom regime  \label{subsection Milgrom regime}}
\subsubsection*{\small Two additional Lagrangian terms}

 In the Milgrom regime a cubic kinetic term  for the scalar field (fractional in the  quadratic kinetic expression) is assumed  like in most modified gravity approaches  aiming at a relativistic generalization of MOND. It is essentially the cubic term to which RAQUAL reduces
  in the deep MOND regime \citep{Milgrom:2014,Famaey/McGaugh:MOND}.\footnote{The original RAQUAL Lagrangian term for the potential $\varphi$ is $(-8 \pi \varkappa)^{-1} a_0^2\, \mathfrak{F}\big(\frac{(\nabla  \varphi)^2}{a_0^2} \big)$ \citep[eg.~(2b)]{Bekenstein/Milgrom:1984}. It acquires its deep MOND form for $\mathfrak{F}(x) \rightarrow x^{\frac{3}{2}}$, which is cubic in $\frac{|\nabla \varphi|}{a_0}=x^{\frac{1}{2}}$. }  
 Here it is brought into a scale covariant form of weight $-4$, in order to get a scale invariant Lagrangian density.\footnote{Be aware that  Weyl geometric  scale invariance  (cf. fn.\ref{fn Weyl scaling})   is different from  Milgrom's scale invariance used in the latter's  discussion of  the  deep MOND regime  \citep{Milgrom:2013,Milgrom:2014}.   } 
It will be called  the {\em cubic term} of the kinetic Lagrangian 
\beqarr L^{(cub)} &=& - \frac{2}{3}\beta \phi^{-2}| D\phi |^3 \, , \qquad |D\phi |= |D_{\nu}\phi D^{\nu}\phi|^{\frac{1}{2}} \nonumber \\
 &\underset{E}{\doteq}& - \frac{2}{3}(\beta^{-1} \phi_0)^{-1} \phi_o^2 \, |\nabla \sigma|^3  \label{eq L-Dphi3}
\eeqarr 
The first line gives the scale invariant expression of the Lagrangian density, the second one its specification in Einstein gauge. The Riemann gauge (Jordan frame)  expression can easily be read off from the scale invariant expression because of $D_{\nu}\underset{R}\doteq \partial_{\nu}$. In this section we assume that the {\em  measuring gauge} is  {\em identical to the Einstein gauge}, at least in sufficient approximation; for a refinement see sec. \ref{subsection FRW regime}. 

The dimensional quantity 
\beq  \beta^{-1}\phi_0=a_0 , \qquad  a_0 \approx 3.9 \cdot 10^{-19}\, s^{-1}  \label{eq a_0}
\eeq
is  a {\em} {\em new constant of nature} of physical dimension $T^{-1}$ ($T$ time), standing in a well known relation
 to the  (empirically determined)  MOND acceleration
 
\[ a_0 \, [c] \approx 1.2  \cdot 10^{-10}\, m s^{-2}  \, . 
\]
It is often expressed as $a_0 \approx \frac{H_0}{6}$ ($H_0$ the Hubble parameter); but this is a (cosmologically) transient characterisation  only. A more principled relation exists  with the ``cosmological constant'' $\Lambda$ (the Einstein gauged quartic potential term of the scalar field), 
\[ \Lambda= 36 \lambda \, a_0^2 \, 
\] with $\lambda$ at the order 1 (see below, eq. (\ref{eq Lambda})). 

$a_0$ will be called the {\em Milgrom constant}; it corresponds to the smallest physically meaningful energy quantity of present physics, 
\[ E_M= a_0\, [\hbar] \approx 2.6 \cdot 10^{-34} \, eV  \qquad \mbox{(Milgrom energy)}.  
\]
The quotient 
\beq \beta  = \frac{\phi_0}{a_0} = \frac{E_P}{E_M} \sim 10^{62} \label{eq beta}
\eeq
 is the ``penultimate''  hierarchy factor between the (reduced) Planck energy $E_P$
 and  $E_M$. It  will be called the {\em cosmological hierarchy factor}.

In order to improve on the RAQUAL approach we have to add an energy carrying contribution to the kinetic Lagrangian of the scalar field, here realized by  the
 second (derivative) order term.  It has been introduced by Novello/Oliveira et al.  in 1993 for  cosmological considerations \citep{Novello/Oliveira_ea:1993} and will be called  the {\em Brazilian term}: 
\beqarr
L^{(braz)}  &=& -\gamma \phi \,D_{\nu}D^{\nu}\phi \nonumber \\
  &\underset{E}{\doteq}& - \gamma (8\pi G)^{-1} \,(\nabla\hspace{-0.1em}(g)_{\nu} \partial^{\nu} \sigma+ \partial_{\nu}\sigma\partial^{\nu}\sigma) \label{eq L-braz}
  \eeqarr
 Novello et al. noticed that  $\mathcal{L}^{(braz)}$  does not increase the order of the scalar field equation because the derivative second order term  $\sqrt{|g|}\,\nabla(g)_{\nu} \partial^{\nu} \sigma =\partial_{\nu}(\sqrt{|g|} \partial^{\nu}\sigma)$ is a divergence. So it can be shifted to a boundary integral and  does not contribute to the  variation  $\delta \phi$. The scalar field equation  remains of order 2.

All in all, the Lagrangian in the Milgrom regime of WdST is 
\beqarr L_M &=& L^{(H)} +  L^{(\phi\, kin)} + L^{(V_4)}+L^{(bar)} \, , \qquad \label{eq Lagrangian Milgrom regime} \\
 \mbox{where }  \qquad L^{(\phi\, kin)} &=& L^{(q-kin)} + L^{(cub)} + L^{(braz)} \, . \nonumber
 \eeqarr
The  building blocks of the kinetic term tare given in  (\ref{eq L q-kin}), (\ref{eq L-Dphi3}), (\ref{eq L-braz})), $ L^{(H)}$ and $L^{(V_4)}$ in (\ref{eq L^H}, \ref{eq L V-4});  $(\mathit{bar})$ designates  baryonic matter. The  parameters  $\alpha$, $\beta, \, \gamma$, $\lambda_4$ are specified below.   The  Lagrangian of the Milgrom regime can be rewritten  in Einstein gauge  (with $\alpha=-6$) as
\beqarr L_M \underset{E}\doteq (8 \pi \varkappa)^{-1}\Big( \frac{R(g)}{2} &-& \gamma\, \nabla(g)_{\nu}\,\partial^{\nu}\sigma    -  (\gamma +3) \, \partial_{\nu} \sigma \partial^{\nu}\sigma   \label{eq L-M in Einstein gauge}  \\
& & \hspace{6em} - \frac{2}{3} a_0^{-1}\, |\nabla \sigma |^3  -  \Lambda\Big)   + L^{(bar)}\, . \nonumber
\eeqarr 
 
In the Milgrom regime WdST  differs from RAQUAL in two respects: (i) the  nonminimal  coupling of $\phi$ to gravity; in the context of IWG it makes the scalar field  a {\em part of the gravitational sector}. (ii) The second order term $ L^{(cub)}$ of the kinetic part of Lagrangian.  (i) implies an indirect coupling of the scalar field to matter in the derivation of the scalar field equation like in JBD, (ii)  endows the scalar field with a non-negligible energy-stress tensor and leads to gravitational light bending in accordance with the inertial dynamics.

Using the notation $\Theta^{(cub)} \underset{E}{\doteq} 8 \pi \varkappa \, T^{(cub)}$, the contribution of $\mathcal{L}^{(cub)} $ 
 to energy-momentum-stress on the right hand side (rhs) of the Einstein equation (see below) can be expressed as (\ref{eq Theta cub})
\[ \Theta^{(cub)}_{\mu\nu} \underset{E}{\doteq} 2 a_0^{-1} (|\nabla \sigma| \partial_{\mu}\sigma\partial_{\nu}\sigma   -\frac{1}{3}|\nabla \sigma|^3  g_{\mu \nu} ) \, . 
\] 
It is  cubic in $|\partial \sigma|$, which makes it {\em negligibly small} in the Milgrom regime.
The contribution of the Brazilian term to the energy-stress tensor of the scalar field is according to eq.~(\ref{eq Theta braz}))
\[ \Theta^{(braz)}\underset{E}\doteq 2 \gamma \, ( { \nabla(g)_{(\mu} \partial_{\nu)}\sigma } - 3\, \partial_{\mu}\sigma\partial_{\nu}\sigma) - \gamma ( { \nabla(g)_{\nu} \partial^{\nu} \sigma } - \partial_{\lambda}\sigma\partial^{\lambda}\sigma) g_{\mu \nu} \, , 
\]
where $\nabla(g)$ denotes the covariant derivative   with regard to the Levi-Civita connection of $g$, and $\nabla(g)^2$ the corresponding d'Alembert operator. 
The second order terms of (\ref{eq Theta braz}) are {\em dominant in the Milgrom regime} and  crucial for the energy content of the scalar field.\footnote{Variation  $\delta g_E$ in the Einstein gauge leads to the same  second order term.}

\subsubsection*{\small Parameter choice}
As mentioned already,   $\beta \sim 10^{62}$ is the hierarchy factor between the constants $E_{Pl}$ and $E_M$ (corresponding to $a_0$).  We see below that for $\alpha=-6$  (and only with this choice) the scalar field equation in Einstein frame  acquires the form of a covariant Milgrom equation generalizing the deep MOND equation. This is the reason for assuming  conformal coupling for the scalar field in the Milgrom regime. Because of the Brazilian term this does not imply a vanishing  trace of the matter tensor.
The coefficient of the Brazilian  kinetic term will be chosen as $\gamma=4$. With this choice the light bending  of the scalar field is consistent with the additional acceleration induced by the scalar field (see  sec. \ref{subsection weak field approximation}).  The coefficient $\lambda_4$  of the quartic potential determines the value of  a cosmological constant  $\Lambda=\frac{\lambda_4}{4} \phi_0^2$ in the Einstein frame. Its order of magnitude is ``extremely small'', but  can  be related to a coefficient  $\lambda \sim 1$  by use  of  the cosmological hierarchy factor (squared):  
\beq \Lambda= \frac{\lambda_4}{4} \phi_0^2 \approx \lambda H_0^2 \approx 36 \lambda \, a_0^2
 \, . \label{eq Lambda}
\eeq 
Then $\Lambda$ has the order of magnitude typical for a cosmological constant ($\lambda = 3 \,\Omega_{\Lambda}$);  in the Milgrom regime it is negligible. \\[-2em]

  \begin{center}
 Parameter choice in the Milgrom regime \\[0.5em] 

\begin{tabular}{| c | c | c | c | c | }
 \hline 
$\alpha$ & $\beta$ & $\gamma$ & $\lambda$ & $\lambda_4$ \\ \hline \\[-1.2em]
$-6$  & $\sim 10^{62}$ & 4  & $\sim 1$ & $\lambda\, (12\, \beta^{-1})^2$\\
\hline
\end{tabular}
 
 \end{center}

\subsubsection*{\small Dynamical equations \label{subsection dynamical equations}}
In  integrable Weyl geometric scalar tensor theory the {\em  scale invariant Einstein equation}  
is of the form
\beqarr G = Ric- \frac{R}{2}g &=& \phi^{-2} T^{(bar)} + \Theta^{(\phi)}   \, , \label{eq IWG Einstein eq} \\
\qquad \mbox{with} \quad  \Theta^{(\phi)} &=& \Theta^{(\phi\, kin)}+ \Theta^{(V)} + \Theta^{(H)} \, .
\eeqarr
 $G,\, Ric, \, R$ are the  Weyl geometric quantities, $\Theta^{(\phi)}$ denotes the total contribution of the scalar field to energy-momentum-stress (depending  on the  gravitational regime).  $\Theta^{(H)}$ arises from   the non-minimally coupling scalar field to the Hilbert term (see appendix,  (\ref{eq Theta H})): 
\[  \Theta_{\mu\nu}^{(H)} = \phi^{-2}\big(D_{(\mu}D_{\nu)}\phi^2 - D_{\lambda}D^{\lambda}\phi^2 \, g_{\mu\nu} \big) \, 
\]
\noindent
It is well known in JBD theory \citep{Capozziello/Faraoni,Fujii/Maeda}. In Weyl geometric gravity \citep{Drechsler/Tann,Tann:Diss}; in Einstein gauge it is\footnote{In flat space it agrees   with an (ad hoc) ``improvement'' of the energy momentum tensor introduced by Callan/Coleman/Jackiw in quantum theory \citep{Callan/Coleman/Jackiw}.}
 \[  \Theta_{\mu\nu}^{(H)} 
\underset{E}\doteq  - 2 \nabla(g)_{(\mu}\partial_{\nu)}\sigma + 2\, \nabla(g)^2 \sigma g_{\mu \nu}\, , \quad \mbox{with}\; g=g_{_E} \, .
 \]  

In the decomposition of the Einstein tensors, $G(g,\varphi)=G(g)+G(\varphi)$, some terms of $G(\varphi)= Ric(\varphi)- \frac{R(\varphi)}{2}g$  cancel with some of the rhs. In Einstein gauge we have \citep[sec. 6.2]{Scholz:2020GRG}
\beq G(\varphi)_{\mu \nu}\underset{E}{\doteq} 2 \, \partial_{\mu}\sigma \partial_{\nu}\sigma - 2\, \nabla(g)_{(\mu}\partial_{\nu)}\sigma + (\partial_{\lambda}\sigma\partial^{\lambda}\sigma + 2\, \nabla\hspace{-0.1em}(g)^2 \sigma ) g_{\mu\nu} \, .
\eeq
With the notation
\beq \Theta(\sigma) := \Theta^{(\phi\, kin)}+ \Theta^{(H)}-G(\varphi) \,  \label{eq Theta sigma net}
\eeq
the Einstein equation   acquires a  more familiar form  (in the Einstein frame):
\beq G(g)= Ric(g) - \frac{R(g)}{2}g\underset{E}{\doteq} 8 \pi \varkappa\, T^{(bar)} + \Theta(\sigma) + \Theta^{(V)} \, , \label{Einstein eq Ef}
\eeq 
Note  that  
$\Theta(\sigma)$ is  a  shorthand (not even scale covariant) for a  an important part of the rhs of the  Einstein equation  (\ref{Einstein eq Ef}) in which the scale connection contribution of the Einstein tensor $G(\varphi)$ has been subtractively included.  The (scale covariant) energy momentum of the scalar field is:
\beq \Theta^{(\phi)} = \Theta^{(\phi\,kin)}+ \Theta^{(H)}+\Theta^{(V)}, \qquad \Theta^{(\phi\,kin)}= \Theta^{(q-kin)}+ \Theta^{(cub)}+\Theta^{braz} \label{eq Theta-phi}
\eeq

For a quartic potential, $V=V_4$,  the energy momentum in the Einstein gauge boils down to a cosmological constant term, $\Theta^{(V)} \underset{E}\doteq - \Lambda\,  g_E $  with $\Lambda$ as in (\ref{eq Lambda}). It is negligible in the Milgrom regime. 
From   (\ref{eq Theta sigma net}) and the calculations  (\ref{eq Theta sigma}) ff.  in the appendix we find
\[ \Theta(\sigma)_{\mu,\nu}\underset{E}{\doteq} 2 \gamma\big( \nabla(g)_{(\mu}\partial_{\nu)} \sigma -\frac{1}{2} \nabla(g)^2 \sigma \, g_{\mu \nu} \big)\; + \; \mathcal{O}_2(\partial \sigma)  \, , 
\]
where  $\mathcal{O}_2(\partial \sigma)$ contains the quadratic and cubic  terms in $\partial \sigma$.\footnote{More in detail:
\beqa \mathcal{O}_2(\partial \sigma)_{\mu\nu} &\underset{E}{\doteq} &  (\alpha-6 \gamma-2) \partial_{\mu}\sigma\partial_{\nu}\sigma  + 2 a_0^{-1}\, |\nabla \sigma|\partial_{\mu}\sigma\partial_{\nu}\sigma\\
& & \qquad - \,\big(\frac{\alpha+2 \gamma+2}{2}\partial_{\lambda}\sigma\partial^{\lambda}\sigma  + \frac{2}{3}\epsilon_{\sigma} a_0^{-1}|\nabla \sigma|^3  \big) g_{\mu\nu}
 \eeqa
with  $\nabla=\nabla(g_E)$, $g=g_E$ and   $\epsilon_{\sigma}$ like in (\ref{eq epsilon sigma}).  
}
 The second order derivative terms of $\Theta^{(H)}$ cancel, the remaining ones 
 derive from  $L^{(braz)}$.\footnote{In a scale covariantly rewritten pure RAQUAL they would not be  present.} 
  For clusters and galaxies $(\partial_x \sigma)^2$ is at least 5 orders of magnitude  smaller than $|\partial_x^2 \sigma|$.  Thus it is reasonable to approximate the rhs in the Milgrom regime  by its second order derivative terms:
\beq  \Theta(\sigma)_{\mu,\nu}\underset{E}{\approx} 2 \gamma \,\big(\nabla(g)_{(\mu}\partial_{\nu)} \sigma -\frac{1}{2} \nabla(g)^2 \sigma \, g_{\mu \nu}\big) \;  \label{eq Theta(sigma) approx}
\eeq
In the following this expression will be assumed as characteristic also for the  non-centrally symmetric case.

 The {\em scalar field equation} in the  Milgrom regime can be calculated like in  \citep{Scholz:2020GRG}.
 Like in JBD theory one subtracts the traced  Einstein equation from the variational equation of $\phi$; in this way the trace of the baryonic matter tensor  enters the dynamical equation of the scalar field without presupposing a Lagrangian coupling of the latter to matter (see app. \ref{appendix Milgrom equation}). With the denotation $\mathcal{M}(\phi)$  for the lhs of the scalar field equation we get the {\em scale covariant Milgrom equation}  
  \beq \mathcal{M}(\phi) 
  = -\frac{1}{2} (\beta^{-1}\phi)\, \phi^{-2}\,\big(  tr T^{(bar)} +4\, L^{(V)}-\phi\, \partial_{\phi} L^{(V)}  \big)\, ,\label{eq scale covariant Milgrom eq}
 \eeq  
 with $\mathcal{M}(\phi)$ the (scale covariant) {\em Milgrom operator} given in general form in the appendix ((\ref{eq Milgrom operator scale covariant})). 
 For $V=V_4$  and the  Einstein gauge the Milgrom operator is simply  
\beq \mathcal{M}_E(\phi) \underset{E}{\doteq} \nabla(g_{_E})_{\lambda} \, (|\nabla \sigma |\, \partial^{\lambda}\sigma ) \, . \label{eq Milgrom operator Einstein gauge}
\eeq 
 For an ideal fluid with matter density $\rho^{(bar)}$ and pressure   $p^{(bar)}$ the scalar field equation acquires the form of a covariant generalization  of the non-linear Poisson equation known from deep MOND  (\ref{eq covariant Milgrom equation}):  
 \beq  \mathcal{M}_E(\phi) 
 \underset{E}{\doteq} (4\pi \varkappa)\, a_0 \, (\rho^{(bar)} - 3 p^{(bar)}) \,   \label{eq Milgrom eq}
 \eeq
The lhs of eq (\ref{eq Milgrom eq}) derives from   $L^{(cub)}$;  it only holds  for $\alpha=-6$. In  the following it  is called   {\em covariant Milgrom equation}  (without the addition ``scale covariant'').

\subsection{\small  Weak field approximation,  centrally symmetric case \label{subsection weak field approximation} }

 In  the following section we  presuppose the Einstein frame and use the equality sign $=$   as an abbreviation for  $\underset{E}{\doteq}$;  similarly $\approx$ is used for approximations in Einstein frame where the context is clear.\\
 
Let us shed a glance at the weak field approximation in  the static centrally symmetric case (for details see appendix \ref{appendix weak field approximation}). 
In  isotropic conformal coordinates with $x_0=t, x_1=r, x_2, x_3$  the Weylian metric is given by 
\[ ds^2 \underset{E}\doteq - A(r)\, dt^2 + B(r)\, \big(dr^2 + r^2 d\Omega^2  \big) \, , \qquad \varphi \underset{E}\doteq d \sigma(r)= \sigma'(r)dr \; ,
\]
with $d\Omega^2= dx_2^2 + (\sin x_2)^2\, dx_3^2 $. 

In the  weak field case 
  $ g \underset{E}\doteq \eta + h$ with the Minkowski metric \\
$\eta= \mathrm{diag}(-1,1,r^2,r^2 \sin^2 x_2)$  and $h= \mathrm{diag}(h_{00},h_{11}, h_{11}r^2,h_{11}r^2 \sin^2 x_2)$ we have 
  $A = 1- h_{00}\, , \; B = 1 + h_{11}$. 
Equality up to first order  in  $h, h', h'', \sigma', \sigma'' $ will be  denoted by $\underset{1}=$.

 The Milgrom equation (\ref{eq Milgrom eq}) in the vacuum becomes
\beq 0=  r \sigma''+(1+ r \frac{A'}{A})\sigma'  \underset{1}=  r \sigma'' + \sigma'  \, . \label{eq Milgrom eq weak field centrally symmetric}
\eeq
It is  solved by 
\beq
 \sigma'= C_1 A^{-\frac{1}{4}}\, r^{-1} \; ; \label{eq sigma' centrally symm exact}
 \eeq  
 up to first order it is 
\beq \sigma'  \underset{1}=     C_1 \, r^{-1}, \qquad \sigma   \underset{1}=   C_1 \ln \frac{r}{r_0}  \, ,\label{eq sigma' centrally symm}
\eeq    
with any value for $r_0$.

An inspection of the first components of the Ricci tensor by the reduced Einstein equation  (appendix \ref{appendix weak field approximation}) shows that 
 the weak field  Riemannian  metric is  
given by
\beq h_{00}= - 2 \Phi_N^{(bar)}\, , \qquad  h_{11} =  2 \Phi_N^{(bar)} + \frac{\gamma}{2} \sigma \label{eq h_00 and h_11}
\eeq 
with $\Phi_N^{(bar)}$ the  Newton potential of  baryonic matter.

 Outside a central mass $M$ this is  
\beq ds^2 \underset{1}= - (1- 2 \frac{M}{r}) dt^2  + \big(1+   2\frac{M}{r} + \frac{\gamma}{2} \sigma \big) \big(dr^2 + r^2 (d\Omega^2  \big)) \, . 
 \label{eq weak field solution centrally symmetric}
\eeq 
  In the light of empirical evidence for MOND the  constant  $C_1$ is
\[ C_1= \sqrt{a_1 M} \, , \qquad \mbox{with} \quad a_1= a_0\, c^{-1} \, ;
\] 
hen
  \beq 
\partial_r \sigma  \underset{1}= \frac{\sqrt{a_1\, M}}{r} \, .  \label{eq C1}
\eeq
We see that  that the first order approximation of the relativistic Milgrom equation is identical to the  deep MOND potential. The relativistic correction for $A$ in (\ref{eq sigma' centrally symm exact}) is 
 \[    (1-h_{00}(r))^{-\frac{1}{4}} \underset{1}= (1- \frac{2M}{r})^{-\frac{1}{4}}\underset{1}= (1 +\frac{M}{2r}) \ < 1+ a_1 r \,  ,
 \] 
and thus negligible in the Milgrom regime. 

\subsubsection*{\small Energy tensor of the scalar field}
We have to distinguished between the   energy-stress tensor of the scalar field $\Theta^{(\phi)}$ (\ref{eq Theta-phi}) on the rhs of the scale invariant Einstein equation (\ref{eq IWG Einstein eq}) and the net energy-stress expression   $\Theta(\sigma)$  (\ref{eq Theta sigma net}) on the  rhs of the  equation 
(\ref{Einstein eq Ef}) (the one with the Riemannian part of the  Einstein tensor $G(g)$ on the lhs). 
The  first one, taken in the  Einstein gauge  $\Theta_E^{(\phi)}$,  represents the physical energy-momentum and stress properties of the scalar field, while the latter is crucial for the calculation of the Riemannian part of the metric. In the Einstein gauge  they are at first order (\ref{eq Theta-phi}, \ref{eq Theta sigma net})
\beqarr
\Theta^{(\phi)}_{\mu\nu} &\underset{1}=& (2 \gamma -2) \nabla_{(\mu}\sigma_{\nu)} +(2- \gamma)\nabla^2 \sigma g_{\mu\nu}) \quad  [ - \Lambda g_{\mu\nu}]  \label{eq Theta phi 1st order} \\
\Theta(\sigma)_{\mu\nu} &\underset{1}=& 2 \gamma \big( \nabla_{(\mu}\sigma_{\nu)} - \frac{1}{2}\nabla^2 \sigma g_{\mu\nu} \big) \quad [ + \Lambda g_{\mu\nu}] \, , \nonumber
\eeqarr
with negligible  cosmological contributions in the Milgrom regime. For the central symmetric case and  $\gamma=4$ the term proportional to $g$ (vacuum energy like)  dominates. With
 \[\nabla^2 \sigma  \underset{1}= \sigma'' + \frac{2}{r}C'  =C_1 r^{-2}
 \]
  the energy tensor   of the scalar field $\Theta^{(\phi)}$  has the form  of a  variable vacuum energy    with energy density falling off quadratically, $(4 \pi \varkappa)^{-1}\,  C_1 \,r^{-2}$, plus  a  superimposed negative  pressure term in the radial direction.  The energy expression $\Theta(\sigma)$  in the ``Riemannianized'' Einstein equation (\ref{Einstein eq Ef}) has a similar character (positive energy density, modified vacuum energy tensor) with the peculiar property that its reduction (by subtracting its half-trace times $g$) leads to a vanishing energy component: 
\beq \big( \Theta(\sigma) - \frac{1}{2}tr\, \Theta(\sigma) g \big)_{00} \underset{1}= 0 \,  \label{eq vanishing Theta sigma red 00}
\eeq
This property holds generally,  independent of  central symmetry; it  is  
 of major importance for the weak field approximation.

\subsubsection*{\small Contribution of the  scalar field to acceleration and  light refraction }
  In the Milgrom regime with  Einstein gauge  the scalar field induces  the  contribution 
\beq a(\phi)^j = \partial^j\hspace{-0.15em} \sigma \, g_{00} \,  \label{eq  a(phi)}
\eeq
to the total acceleration of slow motions   (\ref{eq total acceleration}), (\ref{eq Gamma(varphi)}).
In the weak field case   
\beq a(\phi) \underset{1}= - \nabla \sigma \, ,  \label{eq additional acceleration}
\eeq 
i.e.,  $\sigma$ functions as an {\em acceleration potential} of  the scalar field. 

For a spherical symmetric Riemannian metric parametrized by  isotropic conformal radius 
 the gravitational refraction index $n_{grav}$  is well known   (cf. \citep{Evans_ea:1996})
 \beq n_{grav} = A^{-\frac{1}{2}} B^{\frac{1}{2}} \, . \label{eq refraction index conformal}
 \eeq 
Here we have 
$ n_{grav}= (1- h_{00})^{-\frac{1}{2}}(1+h_{11})^{\frac{1}{2}}\approx 1 + \frac{1}{2} (h_{00}+h_{11})$ and  thus
\beq   n_{grav} \underset{1}=  1 - 2 (\Phi_N^{(bar)}+ \frac{\gamma}{4}\sigma)    \label{eq radial refraction index}
\eeq 
In this sense,  
$\frac{\gamma}{4}\sigma $ functions   as the  {\em light bending potential}    of the scalar field. For $\gamma=4$  
it agrees with the acceleration potential, so that we have good reasons to consider it as the ``scalar field potential'' 
\[ \Phi^{(\phi)} = \sigma \,.
\]
 Expecting a similar agreement  for less symmetrical constellations we choose 
\beq \gamma= 4  \,  \label{eq gamma=4}
\eeq 
as  default value for the parameter.  As mentioned above the energy momentum of the scalar field is then dominated by the vacuum energy like term proportional to $g$. The effect  may thus be described as ``light bending by dark energy'' similar to an  observation studied by Zhang  in \citep{Zhang:2021} in a different context.

\subsection{\small Milgrom approximation of  WdST gravity \label{subsection flat space limit}}
\subsubsection*{\small  A relativistic weak field background for MOND}
In this section we continue to work in the Einstein frame and presuppose a general weak field metric $(g,\varphi)$ in the Milgrom regime of WdST,    $g \underset{E}\doteq \eta + h$ and $\varphi \underset{E}\doteq  d\sigma$,  $h= \mathrm{diag}(h_{00},h_{11},h_{22},h_{33})$  without specialization to central symmetry.

For centrally symmetric constellations eq.~(\ref{eq Milgrom eq weak field centrally symmetric}) indicates   that the first order approximation of the Milgrom equation reduces to the flat space non-linear Poisson equation. This is true  more generally:
\[ \mathcal{M}_E(\phi) = \nabla(g)_{\lambda} \, (|\nabla(g) \sigma |\, \partial^{\lambda}\sigma )  \underset{1}=   \nabla_{\lambda} \, (|\nabla \sigma |\, \partial^{\lambda}\sigma )  \, ,  \qquad g= \eta+h\, ,
\]
with $\nabla = \nabla(\eta)$ the flat space operator. 
Up to first order the scalar field equation of WdST in the Milgrom regime is equivalent to the deep MOND equation in flat space (with pressures added)
\beq
 \nabla_{\lambda} \, (|\nabla \sigma |\, \partial^{\lambda}\sigma )  \underset{1}=   (4\pi \varkappa)\, a_0 \, (\rho^{(bar)} - 3 p^{(bar)}) \, .  \label{eq Milgrom 1st order}
\eeq 
At first order the {\em scale connection part of the Weylian metric}   is  {\em determined  by a flat space equation}. It  decouples from the Riemannian part.

On the other hand, the weak field equation for
the $(00)$-component of the Riemannian contribution of the Weylian metric   (\ref{eq R-00 component weak field equation}), 
\[ R_{00}(g) \underset{1}= - \frac{1}{2} \nabla(\eta)^2 h_{00}=  4 \pi \varkappa \, \rho^{(bar)} + \Theta(\sigma)_{00} - \frac{1}{2}tr\, \Theta(\sigma) g_{00}  \,   ,
\]
is independent of the scalar field, because  by (\ref{eq vanishing Theta sigma red 00}) the rhs reduces to the baryonic source term  (see app. \ref{appendix weak field approximation}):
\beq h_{00} \underset{1}= - 2\Phi_N^{(bar)}  \label{eq h_00 as Newton potential}
\eeq
 The other components $h_{jj} \; (j=1,2,3)$ depend on the specific conditions of the system under study. They  induce {\em relativistic first order  corrections of the flat space metric} of MOND,  important, among others,  for the course of light rays  (see above  for the centrally symmetric case).

\subsubsection*{\small Determining  $\sigma$    from the Newton acceleration}

Given the approximate baryonic matter distribution modelled in flat space, 
the scalar field contribution to the acceleration  $a(\phi)$  can be calculated without any symmetry conditions  in  two steps. At first   the  Newton acceleration of the baryonic matter density $a_N :=a_N^{(bar)}$ has to be calculated, respectively the Newton potential $\Phi_N$. A    straight forward (although  a bit tedious) vector calculus calculation shows  that  for 
\beq a(\phi)=  \sqrt{ a_0 |a_N|} \frac{a_N}{|a_N|} \, . \label{eq Qu-MOND a}
\eeq
the function  $\sigma$ defined by
\beq \nabla \sigma= - a(\phi)  \label{eq Qu-MOND sigma}
\eeq
 satisfies  (\ref{eq Milgrom 1st order}). 
With  other words, once the Newtonian acceleration field $a_N$ is known, $a(\phi)$ can be calculated by the   algebraic field transformation (\ref{eq Qu-MOND a}), and  $\sigma$ is found by   integration (\ref{eq Qu-MOND sigma}).

(\ref{eq Qu-MOND sigma}) can also be written as
\[ \nabla^2 \sigma= \nabla \big(\sqrt{a_0\, |\nabla \Phi_N |}  \big)\frac{\nabla \Phi_N}{|\nabla \Phi_N|} = \nabla \Big( \sqrt{\frac{a_0}{|\nabla \Phi_N|}}\nabla \Phi_N \Big)
\]
 In the  literature a similar idea is used for dealing with the deep MOND case.  In  slightly different guise
this is known under the name ``quasi-linear'' (QUMOND)  approach  
\citep{Milgrom:2010}, \citep[p. 46ff.]{Famaey/McGaugh:MOND}.  In our context it  makes sense under the conditions of the Milgrom regime, i.e., for  $ \nabla \sigma \leq 10 a_0$  (with $l=2$ in the sense of sec.~\ref{subsection 3 regimes and transition} below).

\subsubsection*{\small Flat space (Milgrom) approximation of  weak field WdST }
Let us now turn to the flat space picture of the weak field dynamics in the Milgrom regime.  
 Because of  (\ref{eq Milgrom 1st order}) the flat limit of equation (\ref{eq Milgrom eq})  for  pressure free baryonic matter  is  the  deep MOND equation: 
 \beq
 \nabla_{\lambda} \, (|\nabla \sigma |\, \partial^{\lambda}\sigma )  =   (4\pi \varkappa)\, a_0 \, \rho^{(bar)} \,  \label{eq deep MOnd eq}
\eeq 
The $(00)$-component of the Einstein equation (\ref{Einstein eq Ef}) delivers    Newtonian dynamics in the flat space approximation\footnote{The Riemannian  (Levi-Civita) contribution to the acceleration of low velocity matter particles $a(g)^j  =\Gamma(g)_{00}^j$ depends crucially on  $h_{00}$:
\[ \Gamma(g)_{00}^j = \frac{\partial_j h_{00}}{2(1+h_{jj})} \underset{1}= \frac{\partial_j h_{00}}{2} \underset{1}= -\partial_j \Phi_N^{(bar)}\  
\]
and  can thus be read off from the Newton approximation like in Einstein gravity. } 
which can easily be  read off from the first order dynamics of WdST  (using the Weylian   metric  $(\eta+h, \varphi=d\sigma)$).

The  scalar field potential $\sigma =\Phi^{(\Phi)}$ turns into the deep MOND potential  $\Phi_{d}=\Phi^{(\Phi)}$ of the given matter distribution. 
 At first order approximation the gravitational acceleration of  massive particles  is  composed of the Riemannian contribution which can be written as a  Newtonian acceleration
\[ a_N^j = - \partial_j \Phi_N
\] 
and of the additional acceleration (\ref{eq total acceleration})  due to the scalar field
\[ a (\phi)^j =  \Gamma(\varphi)^j_{00} = - \partial_j \sigma = - \partial_j \Phi^{(\phi)} \, .
\] 
The  total acceleration is then 
\[ a_{tot}  \underset{1}= - \nabla (\Phi_N + \Phi^{(\phi)}) = a_N + a(\phi) \, .
\]
This result is the same as in the ``conformal emergent gravity'' approach (CEG) by Hossenfelder and Mistele \citep{Hossenfelder/Mistele:2018}, although both are derived from different principles. 
With (\ref{eq Qu-MOND a}) we get 
\beq a_{tot}  \underset{1}=  \frac{a_N}{|a_N|} \big(|a_N| + \sqrt{a_0\, |a_N|} \big) \, . \label{eq acceleration flat space limit MOND}
\eeq

Denoting  by $a$  the same acceleration  represented in {\em  flat space} we get 
\[  a= \tilde{\nu} (\frac{|a_N|}{a_0})\,a_N \,, \qquad \mbox{with} \quad   \tilde{\nu}(y)=1+y^{-\frac{1}{2}} \;,  \qquad y= \frac{|a_N|}{a_0} \, . 
 \] 
Of course this holds only under the condition  $a(\phi)=\sqrt{a_o\, |a_N|} \leq 10^{\,l-1}\, a_0$,  respectively $|a_N| \leq 10^{2l-2}a_0$, inherited from the delimitation of the Milgrom regime  (cf sec.\ref{subsection 3 regimes and transition}).  The smooth transition function $h(x;a,b)$ introduced there,  (\ref{eq transition function}), has value $0$ for $x\leq a$ and value $1$ for $x\geq b$. With it we can  express  $a$ more generally as
\beqarr  a &=& \nu (\frac{|a_N|}{a_0})\,a_N \, ,  \label{eq a MOND}   \\ 
\mbox{with}\qquad \nu(y) &=& 1+ h(y^{-1};10^{-2l},10^{2-2l})\,  y^{-\frac{1}{2}} \, .
 \label{eq nu WdST MOND}
\eeqarr
This function hallmarks two well defined  acceleration regimes:\footnote{For $10^{2l}> y > 10^{2l-2}$ the value of $a$ is a formal  interpolation between the  regimes.}
\beq
\nu(y) = \bigg\{ {\mbox{\hspace*{0.5em}} 1 \qquad  \mbox{\;\;\; for}\quad  y \gg 10^{2l} \quad \rightarrow \; a=a_N  \; \mbox {(Newton regime)}\;  
\atop 1+   y^{-\frac{1}{2}}  \quad \mbox{for} \quad y \ll 10^{2l-2} \; \rightarrow a=a_{tot} \;  \mbox{(MOND (WdST))} }  
\eeq
A plausible  value  for $l$ is $l=2$.

Functions  of similar  type are used in  the MOND approach    as ``interpolation functions'' between the deep MOND regime ($|a| \approx \sqrt{a_0\, |a_N|}$) and the Newton regime ($a=a_N$).\footnote{A second interpolation function, usually denoted $\mu(x)$, is used. It has the property $a_N=\mu{(\frac{|a|}{a_0})}a$; see. e.g., \citep[sec. 5.1]{Famaey/McGaugh:MOND}. }
They are understood as expressing possible  quantitative specifications  of ``Milgrom's law'' which  postulates a smooth monotonous transition of gravitational dynamics between the Newton acceleration  and the deep MOND one. Such a $\nu$-function has to satisfy two approximation conditions  \citep[sec. 5.1]{Famaey/McGaugh:MOND}:

\[\nu(y) \approx \bigg\{ {\mbox{\hspace*{-1em}} 1 \qquad  \mbox{\;\;\;\;
 for}\quad  y \gg 1 \; \;\rightarrow \; a \approx a_N  \; \mbox {(Newton regime)}\;  
\atop y^{-\frac{1}{2}} \qquad \;\;\; \mbox{for} \quad y\ll 1 \;\; \rightarrow \; a \approx \sqrt{a_0\, |a_N|}\; \;  \mbox{(deep MOND)} }
\]
In our case both are satisfied; i.e. the present {\em flat space  approximation of WdST}  satisfies Milgrom's law. In analogy to the Newton approximation of Einstein gravity it  will  be called  the {\em Milgrom approximation} of WdST, which is the   MOND variant specified by (\ref{eq nu WdST MOND}) (WdST-MOND).

\subsubsection*{\small Some unique features of WdST-MOND}
The dynamics of WdST MOND  has some unique features among the  MOND models and also some crucial differences to the whole class:

\begin{itemize}
\item[(i)] The whole ``interpolation'' range between the Newton regime and the deep MOND region  
  consists of two parts. The larger one, with roughly  $ |a_N|<  10^{2l-2}\, a_0$ ($l=1$ or $2$),  the ``upper'' one in terms of distance, is determined by the dynamics of WdST in the Milgrom regime, expressed by the interpolation function (\ref{eq nu WdST MOND}). This  interpolation function  is the same as in the study of Hossenfelder/Mistele (\citep{Hossenfelder/Mistele:2018}), in which  2693 data points from 153 galaxies are analysed. The authors show  that the mass disrepancy relation for these data is  extremely well reproduced (``predicted'') by   MOND with this interpolation function. 
 
  \item[(ii)]
  Only in the domain where  $10^{2l-2} a_0 <  |a_N|< 10^{2l} a_0 $ (the ``lower'' part in terms of distance) enters the conventionally chosen transition function $h(x;a,b)$. In usual MOND models the whole interpolation function is conventional (with constraints).
  
  \item[(iii)] In the ordinary MOND approach the deep MOND acceleration 
  may be interpreted in terms of a fictitious Newton potential which would induce the same additional acceleration, in the central symmetric case:
 \beq  \Phi^{(ph)} = \sqrt{a_1 M} \ln r \, . \label{eq Phi phantom}
\eeq 
It is sometimes ascribed to fictional ``phantom  matter'' of  unknown origin and ontological status  with density $\rho^{(ph)}$ in agreement with Newton dynamics, $\nabla^2 \Phi^{(ph)} = 4 \pi \varkappa\, \rho^{(ph)}$   
\citep{Hodson_ea:2020}.   MOND protagonists often postulate   gravitational light deflection to be  in accordance   with the phantom matter density, often even with a ``relativistic'' factor 2, assuming that a general relativistic extension of MOND will justify such an assumption \citep[sec. 8]{Famaey/McGaugh:MOND}.
This is consistent with the present empirical evidence of microlensing; but in classical MOND  the physical reason for such an effect is completely unclear. Here  we have shown that  this expectation is satisfied (sec. \ref{subsection weak field approximation}).\footnote{Similar justifications arise, of course, also in other relativistic extensions of MOND like,  e.g., \citep{{Skordis/Zlsonik:2021}}.}

\item[(iv)] The scalar field has an energy tensor $T^{(\phi)}= (8\pi \varkappa)^{-1}\Theta^{(\phi)}$  with a non-negligible energy density $\rho^{(\phi)}$   and stress/pressure terms (\ref{eq Theta phi 1st order}). It resembles a  type of dark energy  {\em sui generis} which is {\em not} ``phantom''. It  supports the view  that there need not be a strict dichotomy between modified gravity  and dark matter/energy explanations of galactic and cluster dynamics, cf.   \citep{Martens/Lehmkuhl:2020}.

\item[(v)]  The energy density of the scalar field  $\rho^{(\phi)}$ does {\em not} enter the  $(0,0)$-component of the relativistic weak field  approximation  (\ref{eq vanishing Theta sigma red 00}).
In spite of this the scalar field energy-momentum tensor as a whole enhances  gravitational light deflection due to the baryonic matter sources through its pressure components. In the central symmetric case,  the scalar field potential adds up  to the  Newton potential in the calculation of the light refraction index. For $\gamma=4$ the  role of the phantom density for gravitational light deflection in the radial direction even in its relativistic extension (by factor 2) conjectured by some authors of the MOND program,  is  justified   (\ref{eq radial refraction index}). 

\item[(vi)] 
We have seen that the scalar field builds up a halo of energy density around mass concentrations in the Milgrom regime, which can be calculated in the weak field approximation (\ref{subsection flat space limit}). We ´expect a superposition of the halos of ``dark'' energy density in the neighbourhood of single galaxies and the halo  accumulated  at the scale of a galaxy cluster,   calculated in a weak field (MOND) approximation at this larger scale. This makes an important difference to  models of cluster dynamics in the usual MOND  approach. A heuristic evaluation of cluster data on the basis of a  precursor approach to the present one indicates that  no additional dark matter besides the energy tensor of the scalar field   may be necessary  to get the dynamics at   cluster level right  \citep{Scholz:Clusters}. This  conjecture  ought to be checked  by astronomers in more detail.

\end{itemize}

\section{Other regimes \label{section  other regimes}}
\subsection{\small Transition between the regimes and the overall  Lagrangian \label{subsection 3 regimes and transition}}

The  gravitational regimes introduced in sec. \ref{subsection grav regimes 1} can be  distinguished by the gradient of the solution of  eq.~(\ref{eq Milgrom eq}), even in cases where the solution is only  formal. We characterize the delimitation of the regimes  by threshold values 
for the gradient  of scalar field ($y_1>  y_2 > y_3 > y_4 $) as follows:\\[0.5em]

\hspace{1cm}\includegraphics[scale=1.2]{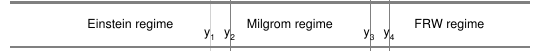}

\noindent 
\begin{itemize}
\item[(1)]{\em Einstein regime} for  $|\nabla \sigma|  \geq y_1$ 
\item[(2)]{\em Milgrom regime} for   $\nabla \sigma$ spacelike  and  $y_2\geq | \nabla \sigma| \geq y_3$
\item[(3)] {\em FRW regime } for $\nabla \sigma$ timelike and/or $|\nabla \sigma|\leq  y_4$ 
\end{itemize}
with roughly  $y_1 = 10^{l+2}\, a_0, \; y_2 = 10^{l} a_0, \, y_3=10^{-k}\, a_0, \, y_4 = 10^{-k'} a_0$, for some $ k<k'$. In the light of cluster dynamics  $l=1$ or $2$ seems reasonable. 
The regions with $|\nabla \sigma|$ between $y_1, \, y_2$ and $y_3, \, y_4$ are {\em transition zones} between the regimes.

\begin{figure}[h]
\centerline{ \includegraphics[scale=0.7]{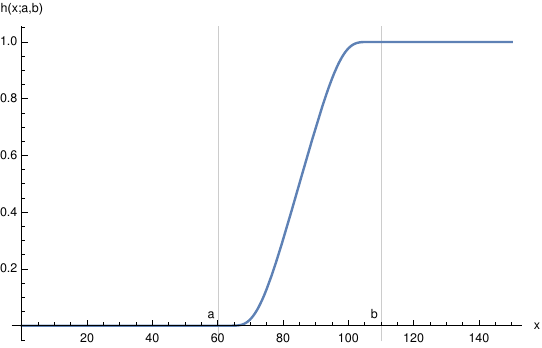} }
\caption{\small Transition function $h(x;a,b)$ \label{fig transition function}}
\end{figure}

For the transition  we use the standard {\em smoothing function} $h(x;a,b)$ 
defined stepwise by
\[ f(x) = 
 \Big\{
{ e^{-\frac{1}{x}} \quad \mbox{for}\quad x>0   
\atop \; \; 0 \qquad \;   \mbox{for} \quad x \leq 0\, , }
\]
from which one builds
\[
g(x) := \frac{ f(x)}{f(x)+  f(1-x)}\, , 
\]
and finally
\beq h(x;a,b) =  g\big(\frac{x-a}{b-a}\big) \, . \label{eq transition function}
\eeq  
It is constant with values $0$  for $x\leq a$  and  $1$ for  $x \geq b$, with a  smooth transition in between (fig. \ref{fig transition function}).

The functions 
\[ h_1(x)= h(|\nabla \sigma(x)|^{-1},y_1^{-1},y_2^{-1} ),\quad h_2(x)= h(|\nabla \sigma(x)|^{-1},y_3^{-1},y_4^{-1} ) 
\]
characterize  the transition between the  Einstein regime (where $|\nabla \sigma(x)| \geq y_1$) with 
$h_1(x)=0$  and  the Milgrom regime + beyond (where $|\nabla \sigma(x)| \leq y_2$)  with   $h_1(x)=1$.  For the transition between the Milgrom regime 
and the FRW regime  $h_2$ plays a similar role. 

The total Lagrangian of the present model is composed of 
the Lagrangians  $L_E$ (\ref{eq Lagrangian Einstein regime}) ,\, $L_M$ (\ref{eq Lagrangian Milgrom regime}) and a hypothetical {\em Ansatz} (\ref{eq Lagrangian FRW regime}) for the cosmological regime $ L_{FRW}$,     with the respective transition functions: 
\beq L = (1-h_1) L_E+ h_1 (1-h_2)  L_M + h_2 L_{FRW} \label{eq total Lagrangian}
\eeq
In Einstein gauge  the  terms of the Lagrangian are given  by   (\ref{eq L-E in Einstein gauge}),   (\ref{eq L-M in Einstein gauge})  and ((\ref{eq L-FRW in Einstein gauge}) respectively.

 The transitions between the regimes  
are represented in the total  Lagrangian (\ref{eq total Lagrangian})  formally, i.e., without making a   claim for   their unknown dynamics.  
The physical reason for the first transition may be a destabilization of the   Milgrom regime scalar field,   if the  gradient  of  $| \nabla \sigma|$ becomes  too strong. This is to be expected if the scalar field represents  a ``superfluid phase''  of some medium  (Koury/Berezhiani, similarly Hossenfelder/Mis\-te\-le);  see sec. \ref{subsection grav regimes 1}. The Lagrangian in the FRW regime considered below (\ref{eq Lagrangian FRW regime}) is a toy model and has  to be improved. 

\subsection{\small Einstein regime \label{subsection Einstein regime}}
If the transition between the Milgrom and  Einstein regimes is due to a destabilization of a superfluid phase, which suppresses the gradient of the physical $\sigma$, the scalar field  becomes constant in the Riemann gauge ($\nabla
 \sigma = 0$), 
 \[ \phi\underset{R}\doteq \phi_0 \, .
 \]
  This implies {\em Riemann gauge = Einstein gauge}. With other words   the integrable Weyl geometry becomes 
  Riemannian and {\em  IWG gravity reduces to Einstein gravity}. This leads to  Einstein gravity written scale covariantly in the geometric framework of IWG, i.e.~basically the gravity theory (WIST) studied by the Brazilian group, although with a different Lagrangian principle behind it.\footnote{The Palatini variation principle used in  WIST implies that the Riemann gauge  is identical with the Einstein gauge \citep[sec. II]{Pucheu_ea:2016}.}

In our approach (WdST) the ``freezing'' of  the scalar field in the Riemann gauge can be be expressed   by a Lagrange multiplier term  with scale covariant  multipliers $\lambda_{\nu}(x)$ (weight $-4$), which fixes the dynamical degree of freedom of the scalar field: 
\[  L^{(\phi, \lambda)}=  \sum_{\nu}(\phi^3 D_{\nu}\phi -\lambda_{\nu})
\]
The Lagrange constraint is  $D_{\nu}\phi=0 \leftrightarrow \partial_\nu \phi \underset{R}\doteq 0$. The scalar field is trivialized (constant in the Riemann gauge) which reduces the Weyl geometric scalar field theory to Einstein gravity. 
The general form of the Lagrangian for Einstein gravity in our framework is
\beq L_E =  L^{(H)} +\sum_{\nu}(\phi^3 D_{\nu}\phi -\lambda_{\nu})+ L^{(m\,bar)} \, . \label{eq Lagrangian Einstein regime}
\eeq
In Einstein (= Riemann) gauge  it reduces to the well known form
\beq
L_E \underset{E}\doteq (16 \pi \varkappa)^{-1} R(g) + L^{(m, bar)} \label{eq L-E in Einstein gauge}
\eeq

\subsection{\small FRW regime and a non-singular cosmological  model \label{subsection FRW regime}}
\subsubsection*{\small No  extension of the Milgrom regime Lagrangian to the FRW context}
We start with  a short look at the aspects of cosmology which can be modelled  by a homogeneous and isotropic spacetime of FRW.  The Riemannian component of the Weylian metric can  be written  in any gauge in a standard FRW form. Here we concentrate on the Einstein gauge with Riemannian component $g_E$:
\beq ds^2 \underset{E}\doteq - dt^2  + a(t)^2 \big( \frac{dr^2}{1-k r^2} + r^2 d\Omega_k^2  \big )
\eeq
(where $d\Omega_k^2$ denotes the line element on the unit sphere of the 3-geometry with constant scalar curvature $k\in{-1,\, 0 , \, +1}$).
If we write  the scalar field in Riemann gauge as $\phi_R \underset{R}\doteq \phi_0 e^{-\sigma(t)}$ like above, the scale connection in Einstein gauge is given by (\ref{eq varphi Einstein gauge 2}), i.e., 
$\varphi_E = \dot \sigma dt$.

The Einstein equation can  be brought into the Friedmann form with contributions $\Theta(\sigma) + \Theta^{(V)}$  of the scalar field in (\ref{Einstein eq Ef})  on the rhs:
\beqarr  \mathit{(i)} \qquad  \big( \frac{\dot{a}}{a} \big)^2 + \frac{k}{a^2} &\underset{E}\doteq& \frac{8 \pi \varkappa}{3} \rho^{(bar)} + \frac{\Theta(\sigma)_{00}}{3} + \frac{\Lambda}{3}  \label{eq Friedmann}\\
 \mathit{(ii)} \hspace{5.5em}  \frac{\ddot{a}}{a}   &\underset{E}\doteq& - \frac{4\pi \varkappa}{3} (\rho^{(bar)}+3 p^{(bar)}) - \frac{\Theta(\sigma)}{6} +  \frac{\Lambda}{3} \nonumber 
\eeqarr 

\begin{figure}[h]
  \includegraphics[scale=0.7]{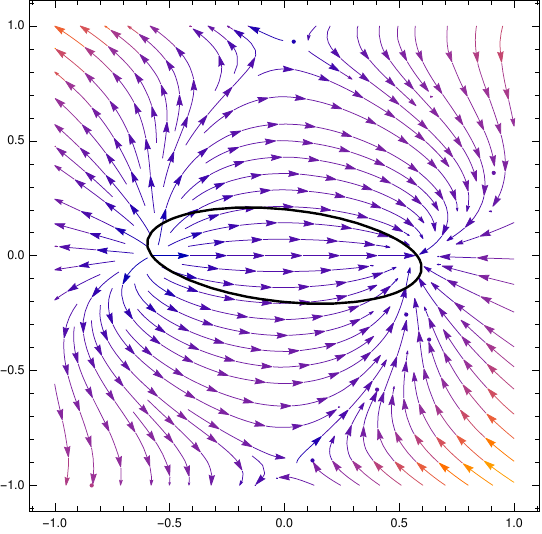} \quad  \includegraphics[scale=0.7]{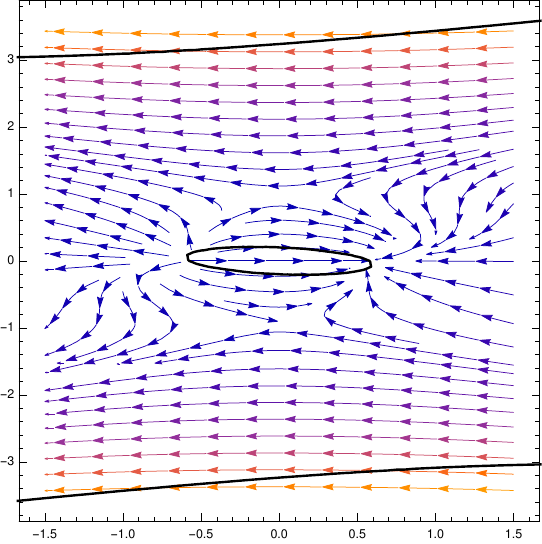}
  \caption{\small Phase diagram for FRW-WdST model  (\ref{eq Lagrangian Milgrom regime}) {  ($\alpha=-6, \, \beta\neq 0, \gamma=4$, $\Lambda=1$)} and algebraic curve (black) transversal to flow lines.  \label{fig phase portrait Milgrom FRW 1}}
   \end{figure}

The Lagrangian  (\ref{eq Lagrangian Milgrom regime})  of the Milgrom regime   with parameters like in section \ref{section WdST}, $\alpha=-6, \, \beta\neq 0, 
\, \gamma=4  $ and $k=0$ is not   compatible with the symmetry conditions of the FRW spacetime. This can be seen by a qualitative investigation of the system of differential equations using the method of   \citep{Kolitch/Eardsley:1995}, which is applicable to scalar tensor theories of different types \citep{Pucheu_ea:2016}, \citep[p. 102ff.]{Faraoni:2004}.
For a non-trivial scalar field, $\dot{\sigma}\neq 0$,   the second Friedmann equation (\ref{eq Friedmann} (ii)) and the scalar field equation (\ref{eq scale covariant Milgrom eq}) can be translated into a system   of two first order  differential equations (a vector field $F$ on $\R^2$) in the  new variables  
$ y= \frac{\dot{a}}{a},\,  z= \dot{\sigma} \,  $.  
The first Friedmann equation turns then into an algebraic (here cubic) relation $R(y,z)=0$. The  phase diagram  of the  system $F$  (fig. \ref{fig phase portrait Milgrom FRW 1}) shows that the flow lines of the present dynamical system are transversal to the  algebraic curve, i.e., the Friedmann equations and the scalar field equation are incompatible. For the trivial case, $\dot{\sigma}\equiv
0$, the Riemann gauge and Einstein gauge coincide and an (anti-) de Sitter - Lanczos model, $a(t)= e^{c_1 t + c_2}$ solves the equations.

     \vspace{1em} 
\begin{figure}[h]
  \includegraphics[scale=0.6]{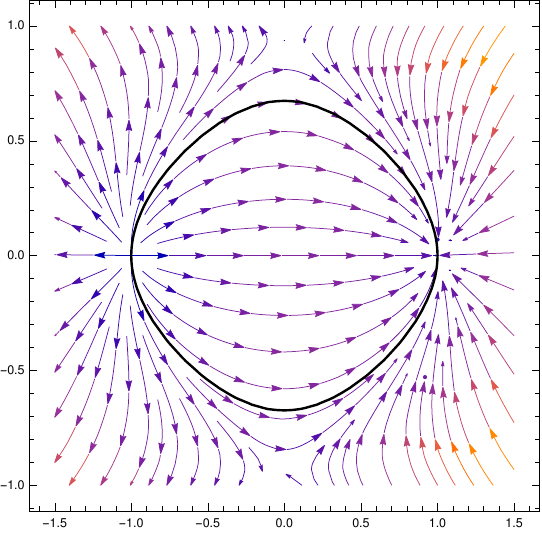} \qquad  \includegraphics[scale=0.8]{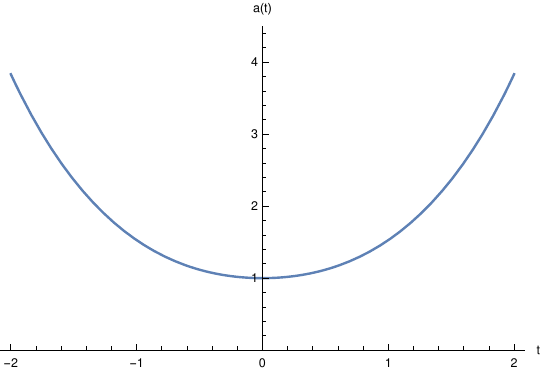}
  \caption{\small  FRW-model  (\ref{eq Lagrangian Milgrom regime}), $\alpha=-6, \, \beta\neq 0, \gamma=0, \,\Lambda=3$. Left: Phase diagram and algebraic curve (black). \quad 
  Right:  (Bouncing) solution for $a(t)$, time unit $H_o^{-1}$. \label{fig phase portrait Milgrom FRW 2}}
   \end{figure}

For $\gamma=0$, however, and  $\alpha=-6, \, \beta\neq 0,  
\,k=0$ the situation is different. Here we find  a flow-line of the vector field along the compact component of the cubic curve (fig. \ref{fig phase portrait Milgrom FRW 2} left). The numerical solution of the original equations satisfies the first Friedmann equation only up to numerical errors,  indicating the existence of a    an exact bouncing solution  of the dynamical equations for $\gamma=0$ (fig. \ref{fig phase portrait Milgrom FRW 2} right). In any case,  a continuation of the  Milgrom regime Lagrangian  to  the cosmological scale does not make sense without a major modification.

\subsubsection*{\small Other conformal (Weyl symmetric)  approaches to  FRW cosmology}

Let us  have a short look at some scale covariant approaches to cosmology,  JBD cosmology \citep{Faraoni:2004}, the Brazilian  version of Weyl geometric  (WIST) cosmology \citep{Pucheu_ea:2016} or, most recently, the ``Weyl symmetric'' cosmological models of Bars, Turok, Steinhart et al. 
with Higgs and gravitational scalar fields \citep{Bars:2014,Bars/Chen_ea:2012,Bars/Steinhardt/Turok:2013,Bars/Steinhardt/Turok:2014}, here called BST cosmology. This is, of course, a small selection from a much wider field (see, e.g., \citep{Cooper/Venturi:1981,Mannheim:1990,Mannheim:2012,%
Ghilencea:2021,Ghilencea/Harko:2021}).

{\em JBD cosmology} has been  studied since its inception in the paper by Brans and Dicke \citep{Brans/Dicke}, extensive  studies of explicit solutions by Lorentz-Petzold \citep{Lorentz-Petzold:1984}, and many other papers. The more recent monographs  \citep{Faraoni:2004,Capozziello/Faraoni} give an impressive  survey of 
 explicit examples and interesting phase space studies for JBD-FRW models. The  explicit solutions for the warp function $a(t)$ and the scalar field $\phi(t)$  discussed there  are often algebraic with fractional exponents and are not promising for realistic cosmological models. On the other hand,  of  Tretyakova and (I.D.)  Novikov et al. have shown that  a realistic looking bouncing solution, including baryonic matter, exists for a negative JBD-parameter $\omega\ll-1$ and a cosmological ``constant'' $\Lambda$ in the Jordan frame \citep{Tretyakova_ea:2012}. 
 
The Brazilian  {\em  WIST approach}  boils down to scale covariantly written Einstein gravity with a minimally coupled scalar field $\phi(x)$. The phase space discussion for cosmological models with different types of potential and the selected explicit solutions studied by Pucheu, Romero et al. \citep{Pucheu_ea:2016}  show   basic features which distinguish WIST cosmology clearly from the JBD  case. From the  perspective of WdST gravity the minimal coupling of the scalar field, however, and the reduction to Einstein gauge = Riemann gauge let  these results appear too special.

That is different for the research program of {\em conformal  cosmology} pursued by 
Bars, Steinhardt, Turok, sometimes with   other authors. In this approach   a consistent  ``Weyl symmetric'' (their terminology) approach  linking up to the standard model of elementary particle physics  (SM) is pursued.  Initially  the authors started  investigating two (or more) non-minimally coupled scalar fields $\phi, \, s$  motivated by a  string or even M-theoretic background   in higher dimensions, which was reduced to its ``shadow'' in 4 dimensions \cite{Bars:2014,Bars/Chen/Turok:2011,Bars/Chen_ea:2012}. Then they turned towards interpreting one of the scalar fields $s$ as a real field expression of a {\em scale covariant} version of the {\em Higgs field} $H$, essentially its ``expectation value''  \citep{Bars:2014,Bars/Steinhardt/Turok:2014},
\beq s(x)=|H(x)|= \big(H^{\dag}(x)H(x)\big)^{\frac{1}{2}}\, ,   \label{eq definition s}
\eeq
where in unitary gauge 
\[
 H(x) = \left( 0 \atop h(x) \right) \, , \qquad  H_0(x) = \left( 0 \atop h_0(x) \right) \,  ,
\]
with real valued $h_0$   characterizing the ground state, and $  h(x)=h_0(x)+\Delta h(x)$. All constituents are  assumed to be  scale covariant of weight $-1$ (here translated into length weights rather than the mass weights preferred in high energy physics). 
Because   the Lagrangian density $\mathcal{L}_{SM}$   is already ``nearly'' scale invariant, with the only exception of the Higgs  mass, this links up nicely with  SM physics, in particular if one assumes a  common biquadratic potential for $\phi$ and $s$
\beq  L^{(V-biq)}= - \frac{\lambda_H}{4}(s^2 - (\omega \phi)^2)^2 \,  \label{eq biquadratic potential}
\eeq 
and adds a quartic self-coupling of $\phi$ 
\beq L^{(V)}=  L^{(V-biq)} + L^{(V_4)} = - \frac{\lambda_H}{4}(s^2 - (\omega \phi)^2)^2  - \frac{\lambda_4}{4} \phi^4 \ . \label{eq potential V}
\eeq
Also other authors have argued that such a modification  leads to a  scale covariant   SM-Lagrangian (weight -4) with scale invariant density  \citep{Shaposhnikov_ea:2009,Meissner/Nicolai}.
  Generalizing to  ``curved spacetime''  this may motivate  to assume 
 a long-range  scalar field  $s=|h|$, a kind of ``real Higgs field''. 
Its ground state value is obtained in the potential minimum,  
 \beq s= h_0(x) = \omega \phi(x) \, . \label{eq real Higgs field}
 \eeq

The ground state $h_0$ and the Higgs mass $m_H $ appearing in the  ``tachyonic'' looking  mass term, 
 \[  \frac{m_H^2}{2}\, H^{\dag}(x)H(x)\;  \quad \mbox{with } \quad m_H^2 = \lambda_H\,(\omega \phi)^2  \;\; \mbox{from (\ref{eq biquadratic potential})}\, ,
 \] 
are both scale covariant of weight -1. In this perspective, the {\em Higgs field gains mass through its coupling to the gravitational scalar field} $\phi$, while the mass of the gauge bosons and the elementary fermions is due to their  coupling to the Higgs field. The coefficients  have to be  chosen such that    in Einstein gauge $h_0(x)\underset{E}\doteq v$, the electroweak (ew) energy scale. This shows that  $\omega$ plays the role of a {\em hierarchy factor} between the ew energy and the (reduced) Planck energy, 
\beq \omega = \frac{v}{E_P} \sim 10^{-16} \, , \label{eq omega}
\eeq
 $\lambda_H$ is the  coefficient between $v$ and  the mass/energy $m_H$ of the  Higgs boson at the electroweak scale
\[ \lambda_H = \frac{m_H}{v} \approx 0.51 \,.
\]

According to their background program, BST  assume  non-minimal coupling not only for  the gravitational scalar field $\phi$ but also for the (real) Higgs field $s$,   
\[ L^{(H)}_{BST}= (\phi^2 - s^2)R\, ,
\]
where $R$ is the   {\em Riemannian} scalar curvature. They therefore have to  
 postulate conformal coupling of both scalar fields to the Hilbert term. 
 In such a  framework of Riemannian based  conformal gravity the authors develop an intriguing study of FRW cosmological models  with 
 explicit solutions for whole model classes (although often with a   potential different from (\ref{eq biquadratic potential})); they establish geodesic completion for most of their models, some of them bouncing, some even ``cyclic'' etc. \citep{Bars:2014,Bars/Chen_ea:2012,Bars/Steinhardt/Turok:2013,Bars/Steinhardt/Turok:2014}. 
 A clue of their study is the considerate change between different gauges. In their approach the Einstein gauge is characterized by $(\phi^2 -s^2) \underset{E}\doteq const = (8 \pi \varkappa)^{-1} $, Higgs gauge by $ s\underset{H}\doteq const = v$, the ew energy scale, etc. 
These investigations are a challenge and an incentive for Weyl geometric studies of cosmology.

 \subsubsection*{\small Weyl's adaptation argument on the measuring gauge reconsidered}
 As SM particles acquire mass by their coupling to the Higgs field their masses are  proportional to  $s(x)$ in this approach. They are  constant in the Higgs gauge only.  For example, the electron with effective mass $ \sqrt{\mu}_e v$ in the SM\footnote{$\mu_e \approx 2.1\cdot 10^{-6}$} 
 has the scale covariant mass $m_e^2= \mu_e  s^2$.\footnote{Take care not to confound  Weyl geometric scaling with the running of mass with the energy scale. \label{fn Weyl scaling vs running constants}}
Assuming a scale invariant fine structure constant $\alpha$,  the Rydberg constant $R_{ryd}=  \frac{\alpha^2\,  c }{4 \pi \hbar}   m_e $ responsible for the spectral frequencies of atoms  turns also into a scale covariant quantity of  weight -1, with its conventional value in the Higgs  gauge:\footnote{Like $m_e$,  the fine structure constant $\alpha$ becomes dependent on the energy scale under field quantization. Here we consider the low energy effective value of $\alpha$   only, and demand its invariance under Weyl rescaling comparable to $c$ and $\hbar$ (cf. footnote \ref{fn Weyl scaling vs running constants}).}
 \begin{equation} R_{ryd} = \frac{\alpha^2 c }{4 \pi \hbar}   \sqrt{\mu_e} s  \underset{H}\doteq \frac{\alpha^2 c }{4 \pi \hbar}  \sqrt{\mu_e} v 
 \,  \label{Rydberg constant I}
\end{equation}
The energy eigenvalues of, e.g., the Balmer series in the hydrogen atom are governed by the Rydberg constant $R_{ryd}$ and scale  with the Higgs field $s$,
\begin{equation} E_n =  - R_{ryd} \frac{1}{n^2} = - \frac{1}{n^2}   \frac{\alpha^2 c }{4 \pi \hbar} \sqrt{\mu_e} \,s , \qquad  n \in \mathbb N \, . \label{Balmer}
\end{equation}

As soon as one considers  a scale covariant setting,  
typical {\em atomic time intervals} (``clocks'')  become proportional to  the reciprocal  expectation value of the Higgs field $s^{-1}=|H|^{-1}$ and  are {\em constant in the Higgs gauge}. This boils down to an adaptation  of atomic clocks to the Higgs field. It is similar to,  but now better founded than,  Weyl's ad hoc claim of an adaptation of clocks to the (Weylian) scalar curvature of spacetime,  proposed during his discussion with Einstein  in 1918 (also repeated, among others, in \citep[p. 298]{Weyl:RZM5}). 
In our context the {\em Higgs gauge} is the one in which the ticking of atomic clocks at low energies is directly expressed; it  is the {\em the measuring gauge} of WdST. As long as the Higgs field and the gravitational scalar field are closely linked  by the potential (\ref{eq biquadratic potential}) the Einstein gauge can just as well be used  as  a reliable approximation to the measuring gauge (see sec. \ref{section WdST}).

\subsubsection*{\small Higgs portal and the gravity sector}
 The assumption of conformal coupling of the (real) Higgs field $s$ is not  compulsory  in WdST; in the light of the standard kinetic term of the SM Higgs field  
it even seems implausible. In the present framework it would be no problem to implement 
a long range real valued scalar field $s$ 
with a standard kinetic term 
$ L^{(s\, kin)} = - \frac{\bar{\alpha}}{2} D_{\nu}sD^{\nu}s  \, $ 
 and $\bar{\alpha}=1$. Here we confine ourselves to the even simpler assumption that $s$ is    a scalar function expressing the {\em expectation value of the Higgs field} (\ref{eq definition s}) without a proper field dynamics at the classical level and thus without a kinetic term of its own.
The  gravitational scalar field $\phi$, on the other hand,   couples to the Higgs sector by the common biquadratic potential (\ref{eq biquadratic potential}). 
 Spoken metaphorically the  gravitational scalar field enters   the {\em Higgs portal}  and connects the gravity sector  with the Higgs field   \citep{Bernal_ea:2019}. 
  
  For the cosmological regime we  consider  the WdST Lagrangian
\beq L_{FRW} = L^{(H)} + L^{(q-kin)}  + L^{(V-biq)}+ L^{V_4} + L^{(m)}\, , \label{eq Lagrangian FRW regime}
\eeq 
with 
\beq \alpha=-6, \;  \beta=\gamma=0 \, . \label{eq parameter FRW regime}
\eeq 
With these parameters the Einstein gauged Lagrangian is 
\beq  L_{FRW} \underset{E}\doteq    (8 \pi \varkappa)^{-1} \big(\frac{R(g)}{2}  -  3 \nabla(g)_{\nu}\partial^{\nu} \sigma -  \Lambda  \big) -\frac{\lambda_H}{4}(s^2 - v^2)^2 + L^{(m)} \, ; \label{eq L-FRW in Einstein gauge}
\eeq 

The ``net'' dynamical equation  for $\phi$ (after subtracting the trace of the Einstein equation)  (\ref{eq net scalar field equation}) reduces  to
\[ \phi \, \partial_{\phi} L^{(V)} - 4 L^{(V)} = tr\, T^{(m)} \, .
\]
The $V_4$ contributions to the potential (\ref{eq potential V}) on the lhs cancel and only the ones derived from the biquadratic term remain. This leads to the algebraic relation
\beq s^2((\omega \phi)^2 -s^2) = - \lambda_H^{-1}\, tr\, T^{(m)} \, . \label{eq phi FRW}
\eeq
In consequence the {\em    conformal coupling} of the scalar field $\phi$  is here {\em compatible} with the presence of a {\em non-vanishing trace of the matter tensor}.  This is a crucial difference to Einstein gravity.
Solving the quadratic equation in $s^2$ with the positive sign we get
\beq 2 s^2 =  (\omega \phi)^2 + \big( (\omega \phi)^4 + 4 \lambda_H\, tr\, T^{(m)}   \big)^{\frac{1}{2}} \, .\label{eq s-square}
\eeq
  
For pressure-less matter with $-tr\, T^{(m)}$ at the order of magnitude of the critical cosmological density $\rho_{crit}=\frac{3 H_0^2}{8 \pi \varkappa}$ we have (in the Higgs gauge) $\rho_{crit} [\hbar^2 c^5] \sim 10^{-11}\, eV \ll v_{ew}^4$. With  (\ref{eq s-square}) this implies
\beq \omega^2 \phi^2 \approx s^2  \,. \label{eq phi FRW vacuum}
\eeq  
Then $s^2$ is in the potential minimum.
{\em In large part of the universe the Higgs gauge and the Einstein gauge are thus  approximately the same.}

 The only exceptions are  regions of extremely high  matter density close to a cosmological singularity.  In such extreme regions  the Higgs  and Einstein gauges may diverge drastically; then the Einstein gauge can no longer be considered as the metrical gauge. This has important consequences for the physics close to a cosmological singularity, which cannot be pursued further at this occasion.\footnote{A divergence between Higgs gauge and Einstein gauge may also arise in the BST approach.}

 \subsubsection*{\small Two toy  models of WdST-cosmology}
With the parameters (\ref{eq parameter FRW regime}) the  rhs contribution to the Einstein equation (\ref{Einstein eq Ef}) is   $\Theta(\sigma)_{00}\underset{E}\doteq  -6\dot{\sigma}^2, \, \Theta(\sigma)_{jj} \underset{E}\doteq  -2 \dot{\sigma}^2 g_{jj}$. For $k=0$  the  Friedmann equations  become:
 \beqarr  \qquad  \big( \frac{\dot{a}}{a} \big)^2  &\underset{E}\doteq& \frac{8 \pi \varkappa}{3} \rho^{(bar)}- 2\dot{\sigma}^2 + \frac{\Lambda}{3} \label{eq Friedmann eq FRW-WdST}  \\
 \hspace{5.5em}  \frac{\ddot{a}}{a}   &\underset{E}\doteq&  -\frac{4 \pi \varkappa}{3} (\rho^{(bar)} + 3p^{(bar)})  + 2 \dot{\sigma}^2 +  \frac{\Lambda}{3} \,  \nonumber 
\eeqarr
The scalar field equation  (\ref{eq phi FRW}) is trivialized by the potential condition (\ref{eq phi FRW vacuum}). 
In the baryonic vacuum this set of equations has a bouncing solution:
\beqarr
a(t) &\underset{E}\doteq&  c_0  \Big( \cosh \big(\sqrt{\frac{\Lambda}{3}} (2 t + c_1)\big) \Big)^{\frac{1}{2}} \,  \label{eq bouncing exact}  \\ 
\mbox{with} \qquad \dot{\sigma}(t)  &\underset{E}\doteq& \sqrt{\frac{\Lambda}{6}} \Big(\cosh  \big(\sqrt{\frac{\Lambda}{3}} (2 t + 3 c_1) \big) \Big)^{-1} \,  \nonumber
\eeqarr
One can check that the scalar field energy density $\Theta^{(\phi)}_{00}$ (\ref{eq Theta-phi}) remains positive ($\geq 0$), although $\Theta(\sigma)$ is negative.
  \begin{figure}[h]
  \includegraphics[scale=0.7]{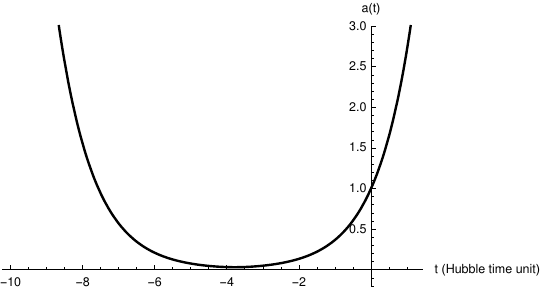} \quad \includegraphics[scale=0.7]{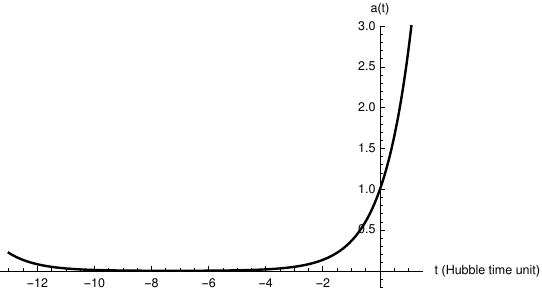} 
  \caption{\small Bouncing solutions (\ref{eq bouncing exact}) ($\Lambda=3$). Left:    at $t_0=0$  redshift $z_{max}=30$   from $t_{min}=-3.78\, H_0^{-1}$. \quad Right:    $z_{max}=1000$ from $t_{min}=-7.26\, H_0^{-1}$.  \label{fig bouncing exact}}
   \end{figure}

If we denote  by $t_0=0$ the present time,  by $t_{min}$  the time of the minimum (``big compression'') 
 $c_0$ and $c_1$ can be chosen such that $a(t_0)=1$ and the maximal redshift observable at $t_0$ gets an arbitrarily prescribed value  $z_{max}$.\footnote{$c_1 = (3 \Lambda)^{-\frac{1}{2}} \arccosh\big( (1+z_{max})^2 \big) , \, c_0=\big(\cosh(\sqrt{3 \Lambda}c_1)\big)^{-\frac{1}{2}}$} 
Clearly this is a   purely formal (``toy'') example.  The deceleration $q=-\frac{\ddot{a} a}{\dot{a}^2}$ in thismodels is  $q(t_0)=-1$,  thus not compatible with the astronomically determined  empirical value $q_0\approx -0.65$.\footnote{For $\Lambda=3$   the Hubble parameter becomes $H(t_0)= \frac{\dot{a}}{a}(0)=1$. Then the time unit is the Hubble time $H_0^{-1}$; for other values of  $\Lambda$ the unit is  changed.} 
 
\begin{figure}[h]
\includegraphics[scale=0.7]{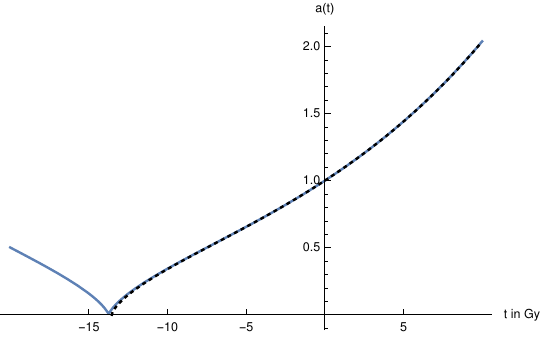} \qquad \includegraphics[scale=0.7]{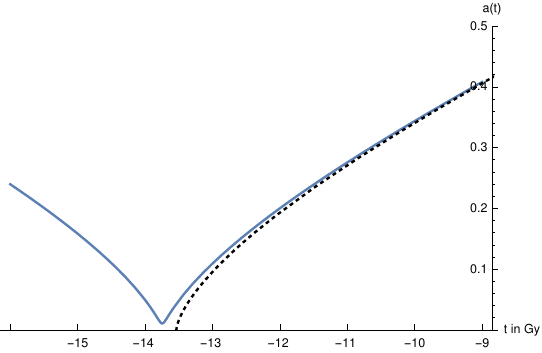}
\caption{\small Blue: Warp (scale) function $a(t)$ in the Einstein gauge  for the FRW-model (\ref{eq Friedmann eq FRW-WdST} with initial conditions (\ref{eq initial conditions FRW-WdsT model}) and parameters $\Omega_m=0.23, \; \Omega_{\Lambda}\approx 0.773 $. Black dotted: expansion for the standard $\Lambda$CDM model with the same parameters. Time in $10^9$ years; the present epoch marked by $t_0=0$, \label{fig FRW-WdST model} }
\end{figure}

An explicit solution of the equations (\ref{eq Friedmann eq FRW-WdST}) with baryonic matter cannot be given as easily. A numerical solution with  initial conditions  at $t_0=0$,   
\beq a(0)=a_0, \;  \dot{a}(0)= H_0\, a_0, \; \ddot{a}(0)= - q_0 H_0^2 a_0 \, , \label{eq initial conditions FRW-WdsT model}
\eeq
 $a_0=1,\,  H_0 = 7.66\cdot 10^{-11}\, [y^{-1}]\; (\sim\; 75\, km\, h^{-1}\, Mpc^{-1}), \; q_0=0.66$ and for realistic parameters expressing the relative energy densities  of matter $\Omega_m$ and of the $\Lambda$-term (arising from the quartic potential $V_4$) $\Omega_{\Lambda}$,
\beq \Omega_m = 0.23, \; \Omega_{\Lambda}\approx 0.773 \, , \label{eq parameters FRW-WdST model}
\eeq 
shows an intriguing   behaviour. It runs extremely close to the present standard $\Lambda$CDM model in the astronomically observable part of the universe until redshift about $z \approx 10$, but then bounces back avoiding a cosmological singularity (fig. \ref{fig FRW-WdST model}). Note  that  we  have $\Omega_m+\Omega_{\Lambda} > 1$ (with the overshoot equal to $2 \,\dot{\sigma}(0)^2 \, H_0^{-2}$).
The time of the minimal warp factor $t_{min}= -13.7\cdot 10^9\,y$ is slightly ``earlier'' than the time of the big bang of the standard model;  the maximal redshift for a present observer is $z_{max}=z(t_{min}) \approx 100$. The energy density of the scalar field  remains positive with the exception of a cosmologically ``small'' time interval about the bounce. 

A squeeze factor $a_0/a(t_{min})\approx 100$  leads to a matter energy density and radiation temperature an order of magnitude below the one necessary for the decoupling of photons ($T=3.6\cdot10^3\,  K$) assumed in $\Lambda$CDM at the surface of last scattering. If we want to explore seriously  whether  this model can serve as an alternative for understanding our real universe, a different origin of the   cosmic microwave background (CMB) has to be found. 
As a first conjecture one might think of  the integrated blue-shifted extra galactic background light emitted at any $t<t_{min}$ and concentrating at the throat ($t_{min}$) (in case the latter thermalizes at the right temperature $273\,  K$).
 But it will be difficult to judge whether the CMB can be understood as a redshifted picture of such a kind of (Olbers) background radiation at the throat.\footnote{A  detailed investigation of the intensity of the  extra galactic background light in  expanding or static, but not in contracting cosmological models can be found in \citep{Wesson:Olbers}. } 
  The answer depends among others,  on a new picture of the formation and passing by of galaxies, independent of the present  one  which has to fight  with the problem of  bringing  the high degree of isotropy of the CMB in agreement with the  the astronomically observed structures already a  ``short''  time after the assumed big bang. 

\section{R\'esum\'e  and discussion  \label{section Outlook}}
\subsection{\small R\'esum\'e \label{subsection discussion}}
In this paper a generalization of Einstein gravity has been studied. It works in  the   scale symmetric framework of integrable Weyl geometry (IWG)  and assumes  a scale covariant scalar field $\phi$ non-minimally coupled to gravity  in addition to baryonic matter. 
At the moment the  most striking effects of this approach can be identified in regions of very weak gravitational fields at the level of galaxies and galaxy clusters  (here called {\em Milgrom regime}). In this regime the gradient of the scalar field adds to the  acceleration of test bodies, while its energy-stress enhances gravitational light deflection (sec. \ref{subsection Milgrom regime}).  Similar to Khoury/Berezhiani's hypothesis of a superfluid theory of dark matter
 (\citep{Berezhiani/Khoury:2015,Berezhiani/Khoury:2016}) the scalar field is assumed to be no longer active in regions where the gradient surpasses a certain threshold; this can be  formally expressed by a Lagrangian multiplier term in the action (sec. \ref{subsection Einstein regime}). Then  the scalar field is inert, i.e., it is  constant in the Riemann gauge,  and the gravitational dynamics reduces to Einstein's theory ({\em Einstein regime}). The Lagrangian of the Milgrom regime  is incompatible with the symmetry assumptions of Friedmann-Robertson-Walker geometry; it  cannot be extended to the {\em FRW-cosmological regime} without change.  In sec. \ref{subsection FRW regime}  a 
a reduction of the Lagrangian in the Milgrom regime  to a conformally coupled scalar field is studied. The latter communicates by a common potential term with a second  scale covariant real-valued scalar field $s$ expressing the expectation value of the Higgs field, 
similar to the proposal by Bars, Turok and Steinhardt  (\citep{Bars/Steinhardt/Turok:2014,Shaposhnikov_ea:2009}). Realistic assumptions for the parameters and initial conditions  lead to a cosmological model with bouncing behaviour. This may shed new light on open questions in present cosmology 
(sec. \ref{subsection FRW regime}). 

 The focus of this paper lies, however, in the Milgrom regime in which central   features of modified Newtonian dynamics  can be derived  on the galactic scale. A strong indication that the ``Keplerian'' laws of galaxy dynamics \citep{Famaey/McGaugh:MOND}) hold for the WdST version of MOND (in the flat space limit) can be found in Hossenfelder/Mistele's study of the mass acceleration-discrepancy \citep{Hossenfelder/Mistele:2018}  in the framework of CEG, which leads to the same interpolation function $\nu$ as ours   (section  \ref{section WdST}).
  
   The present approach shares some characteristics with  Jordan-Brans-Dicke theory in the Einstein frame, but differs in two important respects: the  particle dynamics follows the trajectories of the Weylian metric, not the ones of the Riemannian metric of the Einstein frame (sec. \ref{subsection Differences IWG JBD});  moreover it uses a modified Lagrangian inspired by Bekenstein/Milgrom's RAQUAL  and an additional term  of the Lagrangian with coefficient $\gamma$, first proposed in  a paper by Novello et al. (sec. \ref{subsection Milgrom regime}, eqs.  (\ref{eq L-Dphi3}, \ref{eq L-braz}, \ref{eq Lagrangian Milgrom regime})).  The Einstein equation for the  Riemannian part of the  Weylian metric in the Einstein frame has a  familiar form (\ref{Einstein eq Ef}). The scalar field $\phi$ satisfies a differential equations which, also written down in the Einstein gauge,  is  a  general relativistic generalization of the nonlinear Poisson equation for the deep  MOND potential (\ref{eq Milgrom eq}); therefore it  has been named ``Milgrom equation''. 

 The  contribution  $\Theta^{(\phi)}$ of the scalar field  to the total energy momentum has peculiar properties. Most strikingly, the  $(0,0)$-component of the semi-trace reduced expression $\Theta^{(\phi)}-\frac{1}{2}tr\, \Theta^{(\phi)}\, g$, important for the weak field approximation,  vanishes. This  leads to the  Newton approximation  being sourced by baryonic matter only (\ref{eq vanishing Theta sigma red 00}). As a result the weak field modification $h_{00}$ of the Minkowski metric stands in  the same relation to the Newton potential as in Einstein gravity, $h_{00}=-2 \Phi_N$. In IWG  the trajectories deviate, however,  from the Levi-Civita connection of the weak field (Riemannian) metric; an additional contribution to the  acceleration  (\ref{eq additional acceleration})  comes from the gradient of the scalar field (more precisely from the  gradient of the  ``scalar field potential''  $\sigma$, i.e., the logarithm  of the scalar field in the Riemann gauge $\phi \underset{R}\doteq \phi_0 e^{\sigma}$). For the weak field dynamics of particles we thus encounter a superposition of the Newton potential and the scalar field potential, $\Phi_N + \sigma$.Thus  the particle dynamics is governed by a   superposition of the Newton acceleration and a  contribution induced by the scalar field equal to the  deep MOND acceleration (sec. \ref{subsection weak field approximation}). 
In the flat space limit this leads to  a special case of  MOND  with  ``interpolation function'' $\nu(y)=1+y^{-\frac{1}{2}}$ in the Milgrom regime.

Next it has been shown that for centrally symmetric constellations  the gravitational light refraction stands in agreement with the changed dynamics; i.e., the scalar field potential $\sigma$ also  adds up to the light bending potential of  baryonic matter (\ref{eq radial refraction index}). This is an important result
which distinguishes the present approach from the original RAQUAL proposal of Milgrom and Bekenstein. It depends crucially on the Brazilian term of the Lagrangian and holds only for   $\gamma=4$.\footnote{A similar result  claimed on a flawed basis in \citep{Scholz:2016MONDlike}  is herewith corrected.}

\subsection{\small Discussion and outlook}
The present approach is far from a fundamental theory behind MOND. After all it shows that already one scalar fields suffices for deriving the MOND dynamics in weak field constellations at the level of galaxies and clusters. Integrable Weyl geometry (with the scalar field governing the deviation of the Weyl geometric affine connection in the Einstein gauge  from its Riemannian contribution) signifies an  extremely moderate and  convincing  change of the Riemannian framework of Einstein gravity. No disformal deformation of the metric (TeVeS), no timelike unit field breaking the local Lorentz invariance at the general level (Einstein aether theory), or any other complicated and/or artificial gadget is necessary. All we need is a Weyl symmetric generalization of Riemannian geometry, which is of advantage for high energy physics anyhow and  possibly even an important feature for quantizing gravity \citep{Percacci:RG_flow,Ghilencea:2019JHEP}.  

Like in other   scalar-tensor theories the scalar field  shows interesting qualitative features in addition to its more technical properties. Its  energy-stress tensor $\Theta^{(\phi)} \underset{E}\doteq  8\pi \varkappa\, T^{(\phi)}$ appearing on the rhs of the Einstein equation (\ref{Einstein eq Ef})
 decomposes into two parts, one proportional to the Riemannian  metric $g_E$ like for dark energy,  $\Theta^{(de)} \underset{E}\doteq \tilde{\Lambda}(\phi) g_E$ with  variable coefficient $\tilde{\Lambda}$,  and another one, $\Theta^{(dm)}$, which looks  like a  ``dark fluid''  tensor,
\[
\Theta^{(\phi)} \underset{E}\doteq 
 \Theta^{(dm)} + \Theta^{(de)}   
\] 
This decomposition should, however, not been taken literally. 
For the weak field approximation in the  Milgrom regime  and for $\gamma=4$,e.g.,  we have   found  (\ref{eq Theta phi 1st order})  
\[ \Theta^{(\phi)}_{\mu\nu} \underset{1}= 6\, \nabla_{(\mu}\sigma_{\nu)} -  (2\,\nabla^2 \sigma +\Lambda)\,  g_{\mu\nu} \, ,
\]
where the  ``cosmological'' contribution $\Theta^{(V)}\underset{E}\doteq \Lambda g$ may be neglected. 
In a static weak field approximation this  leads to a positive overall energy density of the scalar field  $\rho^{(\sigma)} \underset{E} \doteq (4 \pi \varkappa)^ {-1} \nabla(g_E)^2\sigma$ which is  due only to the ``dark energy'' contribution $ \Theta^{(de)}$, while the ``dark fluid'' has vanishing energy density.  In the ordinary ``dark'' language, we might describe this as an energy tensor of the scalar field, the most important (energy carrying) part of which has the form of a generalized dark energy tensor  with a superimposed ``dark fluid''-like modification of the pressure components. 
Taken together $\Theta^{(\phi)}$ results in a modification of Einstein gravity with all the effects usually attributed to particle dark matter. In this  sense {\em  the scalar field} $\phi$   plays a double role in the present approach: it {\em  modifies gravity} by its coupling to the Hilbert term and the induced Weylian scale connection in the Einstein gauge,  and it contributes to the rhs of the Einstein equation  like {\em  a peculiar combination of dark matter and dark energy}. In the latter role it underpins the analysis in \citep{Martens/Lehmkuhl:2020} in which an often assumed strict dichotomy between modified gravity and dark matter has been put into question.

Two final remarks on  the cosmological level. (i)  Also in this approach no direct connection between the Milgrom regime dynamics and cosmology could be established. After all an {\em  indirect link} between $a_0$ and $\Lambda$ arises from introducing a common cosmological hierarchy factor $\beta$  ( \ref{eq beta})
and its implementation in the kinetic term of the scalar field (\ref{eq L-Dphi3}) and the scale invariant potential term  (\ref{eq L V-4}, \ref{eq Lambda}).

(ii)    It remains an open question, whether the example of the cosmological model (\ref{eq Friedmann eq FRW-WdST}) (with initial conditions and parameters  (\ref{eq initial conditions FRW-WdsT model}), (\ref{eq parameters FRW-WdST model})) can  develop  into a realistic approach; but  it seems worthwhile to check.  It is not impossible that this, or
 a similar  model built upon a more refined scalar field  approach,  may become a game changer for cosmological modelling 
  leading to more awareness of the astronomically accessible part of the universe than in the present cosmological discourse.  

\vspace*{5em}
\section{Appendix \label{section appendix}}
\subsection{\small Energy-stress  of the scalar field \label{appendix Theta-sigma}}
In  the variation of $\mathfrak{L}^{(H)}$ the non-minimal coupling of the scalar field   results in a term  $\Theta^{(H)}$ (in addition to $Ric-\frac{R}{2}g$)  \citep[eg.~(2.17)]{Drechsler/Tann}, \citep[eg.~(3.5)]{Capozziello/Faraoni}, \citep[sec. 6.2]{Scholz:2020GRG}:
 \beqarr  \Theta_{\mu\nu}^{(H)} &=& \phi^{-2}\big(D_{(\mu}D_{\nu)}\phi^2 - D_{\lambda}D^{\lambda}\phi^2 \, g_{\mu\nu} \big) \nonumber\\
  &=& 2 \phi^{-2}\big(D_{\mu}\phi D_{\nu}\phi + \phi D_{(\mu}D_{\nu)}\phi - (\phi D_{\lambda} D^{\lambda}\phi + D_{\lambda}\phi D^{\lambda}\phi)g_{\mu \nu}   \big) \nonumber \\
&\underset{E}\doteq & - 2 \nabla_{(\mu}\partial_{\nu)}\sigma + 2\, \nabla^2 \sigma g_{\mu \nu} \label{eq Theta H}
 \eeqarr  
  with  $\nabla=\nabla(g_{_E})$ in the whole appendix \ref{appendix Theta-sigma}.

The contribution $\Theta^{(\phi)}$ of the scalar field to the rhs of the Einstein equation (\ref{eq IWG Einstein eq})   derived from the non-Hilbert terms of the Lagrangian contains the kinetic and the potential contributions,
\beq \Theta^{(\phi  \urcorner H)}=  \Theta^{(\phi\, kin)} + \Theta^{(V)}\, , \label{eq Theta phi non-Hilbert}
\eeq
where in the Milgrom regime with the Lagrangian (\ref{eq Lagrangian Milgrom regime})
\beq \Theta^{(\phi \, kin)}=  \Theta^{(q-kin)} + \Theta^{(cub)} + \Theta^{(braz)} \, . \label{eq Theta sigma}
\eeq 
$\Theta^{(X)}$ denotes the contribution of  $X$ to the rhs of the Einstein equation: \[ \Theta^{(X)}=\phi^{-2}T^{(X)} = - \phi^{-2}\frac{2}{\sqrt{|g|}}\frac{\delta \mathfrak{L}^{(X)}}{\delta g^{\cdot \cdot}}
\]

\beqarr
\Theta^{(q-kin)}_{\mu\nu} &=&  \phi^{-2} (\alpha \ D_{\mu}\phi D_{ \nu}\phi+ L_{D\phi^2}\, g _{\mu \nu}  ) \label{eq Theta Dphi2} \\
 &\underset{E}\doteq & \alpha ( \partial_{\mu} \sigma \partial_{\nu} \sigma - \frac{1}{2}\partial_{\lambda}\sigma \partial^{\lambda} \sigma  \, g_{\mu\nu} ) \nonumber \\
 \Theta^{(cub)}_{\mu\nu} &=& 2 \beta  \phi^{-4}\, |D\phi| D_{\mu}\phi D_{\nu}\phi + \phi^{-2 }L^{(braz)}\, g_{\mu \nu} \qquad  \label{eq Theta cub} \\ 
  & \underset{E}\doteq & 2 \beta  \phi_0^{-1}\left( |\nabla \sigma |\partial_{\mu} \sigma \partial_{\nu} \sigma - \frac{\epsilon_{\sigma}}{3} |\nabla \sigma|^3\, g_{\mu \nu} \right) \nonumber \\
   \Theta^{(braz)}_{\mu\nu} &=&  2 \gamma \phi^{-1} \left(- D_{(\mu}D_{\nu)} \phi  + \frac{1}{2}D_{\lambda}D^{\lambda}\phi \, g_{\mu\nu} \right) \label{eq Theta braz} \\
    & \underset{E}\doteq &  2 \gamma \left(\nabla_{\hspace{-0.15em}(\mu}\partial_{\nu)} \sigma  -3\,  \partial_{\mu} \sigma \partial_{\nu} \sigma \right)  - \gamma \left(\nabla_{\hspace{-0.25em}\lambda} \partial^{\lambda}\sigma - \partial_{\lambda} \sigma \, \partial^{\lambda} \sigma  \right) \, g_{\mu\nu} \nonumber \\ 
   \Theta^{(V)}_{\mu\nu} &=&     \phi^{-2}L^{(V)}\, g_{\mu\nu} 
   \eeqarr
where
  \beq \epsilon_{\sigma}= 
 \Big\{
{\mbox{\hspace{-3.5em}} +1 \quad \mbox{for}\quad \nabla \sigma \; \mbox{spacelike}  
\atop \; -1 \quad \mbox{for}\quad\nabla \sigma \; \mbox{timelike or null.} } \label{eq epsilon sigma}
\eeq    
For $V=V_4$    a cosmological constant arises in Einstein gauge (\ref{eq Lambda}):
  \beq    \Theta^{(V_4)}_{\mu\nu} \underset{E}\doteq   - \frac{\lambda_4}{4}\phi_0^2\, g_{\mu \nu} = - \Lambda\, g_{\mu \nu}\, , \;\; \; \Lambda= \lambda\, H_0^2\,  \label{eq Theta V4}
  \eeq 
\noindent
Tracing the (scale invariant) Einstein equation and multiplying with $ - \phi^2$ leads to 
\beqarr && - 2 L_H - tr\, T^{(bar)} - (2 \gamma +6)\, \phi D_{\lambda}D^{\lambda}\phi  + (\alpha +6)  \, D_{\lambda}\phi D^{\lambda}\phi \label{eq trace Einstein eq}  \\
 & & \hspace{17em}  - L^{(cub)} - 4 L^{(V)} = 0 \, . \nonumber
\eeqarr 

\subsection{\small The scalar field equation \label{appendix Milgrom equation}}

  The scale covariant variation with regard to $\phi$, $  \frac{\delta L}{\delta  \phi} =
\frac{\partial L}{\partial  \phi} - D_{\lambda} \frac{\partial  L}{\partial(D_{\lambda}  \phi)} $,
contains the partial contributions \citep[app. 6.2]{Scholz:2020GRG}: 
\beqa      \\
\frac{\delta L^{(q-kin)}}{\delta  \phi} &=& \alpha  D_{\lambda}D^{\lambda}\phi\, \qquad \\ 
\frac{\delta L^{(cub)}}{\delta  \phi} &=& 2 \beta \,\phi^{-2}\, D_{\lambda}\left(|D \phi | D^{\lambda}\phi\right) + 4 \phi^{-1}\, L^{(braz)} 
\eeqa
In the second line we encounter  a scale covariant form of the  non-linear modification of the d'Alembert operator typical for relativistic MOND theories.  

For  $ L^{(braz)}$ it is recommendable to use the Einstein gauge. 
Because of 
\[ \nabla(g)_{\hspace{-0.05em}\lambda}\,  \partial^{\lambda}\, \sigma = \frac{1}{\sqrt{|g|}} \partial_{\lambda} (\sqrt{|g|} \partial^{\lambda} \sigma)
\] 
the second order derivative term of $ \mathfrak{L}^{(braz)}$ in Einstein gauge is a divergence 
\[- \frac{\gamma}{4} \phi_0^2\, \partial_{\lambda} (\sqrt{|g|} \partial^{\lambda} \sigma) \, .
\] 
For the variation of $\phi$ (with fixed $g$) its integral 
 can be shifted to a boundary term outside the support of  $\delta \phi$  and  does not contribute to the Euler-Lagrange equation of the scalar field.\footnote{This has been noted  by the authors of \citep{Novello/Oliveira_ea:1993}.}
This not the case for the variation   $\delta g$.   
 For the variation $\delta \phi$  in Einstein gauge only the  term $-\frac{\gamma}{4}  \phi_0^2\, \partial_{\lambda} \sigma \partial^{\lambda} \sigma $  remains as the reduces Brazilian term.  Its scale covariant version  
 \[ L^{braz\, red}= - \frac{\gamma}{4}  \phi_0^2 D_{\lambda} \sigma D^{\lambda} \sigma\, 
 \]
has the same form as $L^{(q-kin)}$ and leads to a second degree dynamical equation for $\phi$ (respectively $\sigma$). In terms of $\phi$: 
 \beq
\frac{\delta L^{(braz)}}{\delta  \phi} = 2\gamma D_{\lambda}D^{\lambda}\phi \, .
\eeq 
$L^{(H)}$ and $L^{(V_4)}$ are monomials in $\phi$ with $\frac{\delta \phi^k}{\delta \phi} = \frac{\partial \phi^k}{\partial \phi}= k \phi^{k-1}$. \\[0.3em]

After summing up and multiplying with $\phi$ we arrive at the {\em gross  scalar field equation}:
\beqarr 2 L_H + (\alpha - 2 \gamma)\,  \phi\, D_{\lambda} D^{\lambda}\phi + 4 L^{(braz)} + 2 \beta  \phi^{-1}\, D_{\lambda}\left(|D \phi | D^{\lambda}\phi\right) + \phi\, \partial_{\phi} L^{(V)} = 0
\eeqarr
Addition of the traced Einstein equation  (\ref{eq trace Einstein eq}) leads to the scale covariant, (net) {\em  scalar field equation} (in arbitrary gauge):
\beqarr 2 \beta  \phi^{-1}\, D_{\lambda}\left(|D \phi | D^{\lambda}\phi\right) &+& (\alpha + 6) \big( D_{\lambda}\phi D^{\lambda}\phi +    \phi\, D_{\lambda} D^{\lambda}\phi \big) + \nonumber \\
& & \hspace*{-1em} L^{(braz)} +   \phi\, \partial_{\phi} L^{(V)} - 4 L^{(V)} = tr\, T^{(bar)}  \label{eq net scalar field equation}
\eeqarr 
For $\beta=\gamma=0$ and vanishing or quartic  potential,  $V=0$ or $V_4$,  this implies the well known constraint  $ tr\, T^{(bar)} =0$ for conformal coupling ($\alpha=-6$), not so however for different potentials, e.g. the biquadratic one used in sec. \ref{subsection FRW regime}. 

In the {\em MG regime} ($\alpha = -6$) (\ref{eq net scalar field equation})simplifies  to
\[  2 \beta \phi^{-1}\, D_{\lambda}\left(|D \phi | D^{\lambda}\phi\right) + 3 L^{(braz)} =  tr\, T^{(bar)} + 4 L^{(V)}- \phi\, \partial_{\phi} L^{(V)}  \,  .
\] 
Multiplying by $-\frac{1}{2}\, (\beta^{-1}\phi)  \phi^{-2} $ implies   the  {\em scale covariant Milgrom equation} of the main text (\ref{eq scale covariant Milgrom eq}),
\[ \mathcal{M}(\phi)= - \frac{1}{2} \phi^{-2}\, (\beta^{-1}\phi)\, \big( tr\, T^{(bar)}  + 4 L^{(V)}- \phi\, \partial_{\phi} L^{(V)}  \big) \; , 
\]
with  the  the {\em scale covariant Milgrom operator} on the lhs:
\beq \mathcal{M}(\phi)= -\Big( \phi^{-2}\, D_{\lambda}\left(|D \phi | D^{\lambda}\phi\right) -\phi^{-3}|D \phi | \Big) \, \label{eq Milgrom operator scale covariant}
\eeq

In the Einstein gauge we find  $D_{\lambda}(|D\phi |D^{\lambda}\phi) \underset{E} \doteq  -\phi_0^2 D_{\lambda}(|\nabla \sigma|\partial^{\lambda}\sigma)$ and  $\phi \underset{E} \doteq \phi_0$. For a perfect fluid (energy density $\rho_{bar}$ and pressure $p_{bar}$)  and $V=V_4$:
\[  D_{\lambda} \left(|\nabla \sigma | \partial^{\lambda} \sigma  \right) - \epsilon_{\sigma} |\nabla \sigma |^3 \underset{E} \doteq   \frac{1}{2}\phi_0^{-2} \left(\beta^{-1}\phi_0 \right)\, (\rho^{(bar)} - 3 p^{(bar)}) \, .
\]
 The cubic term on the rhs cancels against the one in\footnote{ $|\nabla \sigma|= \epsilon_{\sigma} \partial_{\sigma}\partial^{\sigma}$  and therefore $w(|\nabla \sigma|)=-2$, so  we get
 \beqa D_{\lambda}(|\nabla \sigma| \partial^{\lambda}\sigma) &=&\, _g\hspace{-0.15em} \nabla_{\hspace{-0.15em}\lambda}(|\nabla \sigma| \partial^{\lambda}\sigma) + \, _g\hspace{-0.15em}\Gamma_{\lambda \nu}^{\lambda}|\nabla \sigma| \partial^{\nu}\sigma - 3 \varphi_{\lambda}\, |\nabla \sigma| \partial^{\lambda}\sigma \nonumber \\
 &=& \, _g\hspace{-0.15em} \nabla_{\hspace{-0.15em}\lambda}(|\nabla \sigma| \partial^{\lambda}\sigma) +  \varphi_{\lambda}\, |\nabla \sigma| \partial^{\lambda}\sigma 
 \eeqa
 }
 
 \[   D_{\lambda} \left(|\nabla \sigma | \partial^{\lambda} \sigma  \right)   \underset{E} \doteq \, 
 _g\hspace{-0.2em} \nabla_{\hspace{-0.15em}\lambda} \left(|\nabla \sigma | \partial^{\lambda} \sigma  \right)+ |\nabla \sigma| \partial_{\lambda} \sigma  \partial^{\lambda} \sigma \,
\] 
and the scalar field equation in the Milgrom regime  simplifies to
\beq
\nabla(g)_{\lambda} \left(|\nabla \sigma | \partial^{\lambda} \sigma  \right) \underset{E} \doteq  \quad   4 \pi \varkappa_N \, a_0  \, (\rho - 3p)^{(bar)} \,, \label{eq covariant Milgrom equation}
\eeq
with the Einstein gauged 
 {\em covariant Milgrom operator} for  a scalar field $X$  
\beq    \mathcal{M}_E(X) \underset{E} \doteq   \,  \nabla(g_{_E})_\lambda \left(|\nabla X |\, \partial^{\lambda} X  \right) \,  \label{eq covariant Milgrom op}
\eeq 
on the lhs. 
For the flat metric and static fields  this is the non-linear Laplace operator of classical MOND theory $\nabla_j (|\nabla X|\, \partial^j X)$ (with the Euclidean $\nabla$ operator).

\subsection{\small The weak field approximation for $g$ in the Milgrom regime \label{appendix weak field approximation}}
The  weak field approximation in the  Milgrom regime considers a metric  $ g \underset{E}\doteq \eta + h$ 
where only first order terms in  $h, h', h'', \sigma', \sigma'' $ are considered. The equality up to first order  is denoted by $\underset{1}=$.

Here we concentrate on  the (quasi-)static central symmetric case with conformal spherical coordinates and Weylian metric 
\[ ds^2 \underset{E}\doteq - A(r)\, dt^2 + B(r)\, \big(dr^2 + r^2 d\Omega^2  \big) \, , \qquad \varphi \underset{E}\doteq d \sigma(r)= \sigma'(r)dr \; .
\]
Other cases like the case of cylindrical symmetry can be treated similarly. 
Here we have $\eta= \mathrm{diag}(-1,1,r^2,r^2 \sin^2 x_2), \; h= \mathrm{diag}(h_{00},h_{11}, h_{11}r^2,h_{11}r^2 \sin^2 x_2)$
and  $A = 1- h_{00}\, , \; B = 1 + h_{11}$, where ``$=$'' stands for $\underset{E}\doteq$. 

\noindent  Important are the following  Levi-Civita connection coefficients   
\beqa \Gamma(g)^1_{00} &=& \frac{A'}{2B} \underset{1}=-\frac{h_{00}'}{2}\, ,  \qquad \Gamma(g)^1_{11}= \frac{B'}{2B} \underset{1}=\frac{h_{11}'}{2} \, , \quad \\
\Gamma(g)^1_{22} &=& -r (1+r \frac{B'}{2B}) \underset{1}=-r(1+\frac{r}{2}h_{11}') \; ,
\eeqa 
the d'Alembert operators
\beqa \nabla(g)^2 f &=& \frac{f''}{B} + \big( \frac{A'}{2AB} + \frac{2}{rB} + \frac{B'}{2B^2} \big)f' \underset{1}= f'' + \big( - \frac{h_{00}'}{2} + \frac{h_{11'}}{2} + \frac{2}{r}  \big) f' \\
\nabla(\eta)^2 f  &\underset{1}=&  f'' + \frac{2}{r}f' \; ,
\eeqa
and the  Ricci tensor components:
\beqa R_{00}(g) &=& - \frac{A'^2}{4AB}+ \frac{A'B'}{4B^2} 
 + \frac{A'}{rB}+ \frac{A''}{2B^2} \; \underset{1}= \; - \frac{1}{2}(h_{00}''+ \frac{2h_{00}'}{r}) \; \underset{1}= \; - \frac{1}{2}\nabla(\eta)^2 h_{00} \\
 R_{11}(g) &=&  \frac{A'^2}{4A^2}+ \frac{A'B'}{4AB}- \frac{A''}{2A}+ \frac{B'^2}{B^2}- \frac{B'}{rB} - \frac{B''}{B^2}
\; \underset{1}= \; \frac{h_{00}''}{2}- \frac{h_{11}'}{r}-h_{11}'' \\
 R_{22}(g) &=& -r\, \big( \frac{A'}{2A}+ \frac{r A'B'}{4AB} - \frac{r B'^2}{4B^2}+ \frac{3B'}{2B}+ \frac{r B''}{2AB}      \big)
\; \underset{1}= \; \big(   \frac{h_{00'}}{2r}- \frac{3h_{11}'}{2r} - \frac{h_{11}'}{2} \big) r^2 \\
\eeqa
Consider the half-trace (times $g$) 
{\em reduced Einstein equation}\footnote{The contribution of $\Theta^{(V_4)}$ is cosmologically small ($\Lambda g$) and therefore  negligible.}
\[ Ric(g) \underset{E}\doteq (8 \pi \varkappa) \, \big( T^{(bar)}- \frac{1}{2}tr\, T^{(bar)}\, g  \big) + \Theta (\sigma) - \frac{1}{2} tr\, \Theta(\sigma)\, g \; .
\]
The terms on the rhs will be called the {\em reduced} energy tensors of baryonic matter and of the scalar field:
\[ T^{(b,\, red)}=T^{(bar)} - \frac{1}{2}T^{(bar)} \, g                                                                                                                                                                                                                                                                                                                                                                                                      \;, \qquad  \Theta(\sigma)^{(red)}= \Theta(\sigma) -  \frac{1}{2} tr\, \Theta(\sigma)\, g \,  \]
For  pressure free baryonic matter  with $T_{00}^{(bar)}=\rho^{(bar)}$    the 
reduced energy tensor has components $T_{00}^{(red)}=\frac{1}{2}\rho^{(bar)}$ and  $T_{jj}^{(bar)}=\frac{1}{2} \rho^{(bar)}g_{jj}$.
For the scalar field the (semi-trace) reduction of (\ref{eq Theta(sigma) approx}) leads to 
\beq \Theta(\sigma)^{(red)}_{\mu\nu} \underset{1}= 2 \gamma \, \nabla(\eta)_{(\mu}\partial_{\nu)}\sigma  \; , \label{eq Theta-sigma red}
\eeq
with  components
\[ \Theta(\sigma)^{(red)}_{00 } \underset{1}= 0 \, , \qquad \Theta(\sigma)^{(red)}_{11} \underset{1}= 2 \gamma \sigma'' \, , \qquad  \Theta(\sigma)^{(red)}_{22}   \underset{1}= 2 \gamma \sigma r \; .
\]
This shows that the {\em scalar field does not contribute to the energy component of the reduced Einstein equation}. Note that (\ref{eq Theta-sigma red}) is independent of the assumption of central symmetry and  the energy component of $\Theta(\sigma)$ vanishes in the static case.

We therefore get
\beq R_{00}(g) \underset{1}= - \frac{1}{2} \nabla(\eta)^2 h_{00}=  4 \pi \varkappa \, \rho^{(bar)} \, , \label{eq R-00 component weak field equation}
\eeq
 exactly like in Einstein gravity independent of central symmetry.  
The usual  identification  
\beq  h_{00} = - 2 \Phi_N^{(bar)} \label{eq h_00}
\eeq
leads then to the well known Newton approximation.

 Remember, however,  that in our framework the point particles do not follow the  Levi-Civita connection of $g$  (approximated by the Newton acceleration of baryonic matter),  but are subject to an additional acceleration derived from the scalar field potential $\sigma$ (see sec. \ref{subsection acceleration}).

In the $(11)$ and $(22)$ components of the reduced Einstein equation the scalar field becomes  visible:  
\beqa
R_{11} &\underset{1}=&  \frac{h_{00}''}{2}- \frac{h_{11}'}{r}-h_{11}''  \underset{1}=  4\pi \varkappa \, \rho^{(bar)} + 2 \gamma\sigma'' \\
R_{22} &\underset{1}=&   \big(\frac{h_{00'}}{2r}- \frac{3h_{11}'}{2r} - \frac{h_{11}'}{2} \big) r^2        \underset{1}= \big(4\pi \varkappa \, \rho^{(bar)} +  2 \gamma \frac{\sigma'}{r}\big) r^2
\eeqa
This leads to
\[
R_{11}(g) + \frac{2}{r^2}R_{22}(g) \; \underset{1}= \; \frac{1}{2} \nabla(\eta)^2h_{00} - 2 \nabla(\eta)^2 h_{11} \; \underset{1}=\; 12\pi \varkappa\, \rho^{(bar)} + 2 \gamma \nabla(\eta)^2 \sigma \, ,
\]
from which
\[ 2 \nabla(\eta)^2 h_{11} \underset{1}= - 4\, \nabla(\eta)^2 \Phi_N^{(bar)} - 2 \gamma\nabla(\eta)^2\sigma \, 
\]
 and finally
 \beq  h_{11}  \underset{1}= -2 \,\big(\Phi_N^{(bar)} + \frac{\gamma}{2}\sigma\big) \, . \label{eq h_11}
 \eeq 
With $\gamma=4$ the gravitational refraction is identical to the relativistic refraction induced by a (Newtonian) potential $\Phi_N^{(bar)}+ \sigma$ and thus in agreement with the acceleration of test particles (see sec. \ref{subsection weak field approximation}).

The weak field approximation of the Riemannian metric in the central symmetric vacuum case with central mass $M$ is:
\beq ds^2 \underset{1}= - (1-\frac{2M}{r}) dt^2  + \big(1+ \frac{2M}{r} - 4 \sqrt{a_1 M}\log \frac{r}{r_0}  \big) \big(dr^2 + r^2 (d\Omega^2  \big)) \, 
 \label{eq modified Schwarzschild metric}
\eeq 
It fits well to the Schwarzschild metric in the Einstein regime and can be glued to the latter by the smooth transition function (\ref{eq transition function}).

\vspace{5em}
\noindent
{\bf Acknowledgements}:\\
\small  The preparation of this paper has profited from  the discussions in the subgroup  A3 of the research group ``Epistemology of the LHC'' financed by the German Research Foundation (DFG). The questions posed by Dennis Lehmkuhl and Niels Martens and  Dennis' long standing scientific and historical interests in Weyliana and Jordan-Brans-Dicke theory have helped  a lot for  keeping me on track of the present research. The publication has been supported by the DFG, grant FOR 2063.

\end{document}